\documentclass[sigconf]{acmart}
\usepackage{acmart-taps}
\usepackage{multirow}
\usepackage{array}

\AtBeginDocument{%
  \providecommand\BibTeX{{%
    \normalfont B\kern-0.5em{\scshape i\kern-0.25em b}\kern-0.8em\TeX}}}

\setcopyright{acmcopyright}
\copyrightyear{2023}
\acmYear{2023}
\setcopyright{acmlicensed}\acmConference[CHI '23]{Proceedings of the 2023 CHI Conference on Human Factors in Computing Systems}{April 23--28, 2023}{Hamburg, Germany}
\acmBooktitle{Proceedings of the 2023 CHI Conference on Human Factors in Computing Systems (CHI '23), April 23--28, 2023, Hamburg, Germany}
\acmPrice{15.00}
\acmDOI{10.1145/3544548.3580652}
\acmISBN{978-1-4503-9421-5/23/04}

\usepackage{color}

\newcommand*{\rv}{\textcolor{black}}

\begin{document}

\title[Designerly Understanding: Information Needs for Model Transparency to Support Design Ideation]{Designerly Understanding: Information Needs for Model Transparency to Support Design Ideation for AI-Powered User Experience}

%\title{Designerly Understanding: Transparency and Information Needs for AI Design Ideation}


\author{Q. Vera Liao}
 \email{veraliao@microsoft.com}
\affiliation{%
  \institution{Microsoft Research}
  \city{Montreal}
  \country{Canada}
}

\author{Hariharan Subramonyam}
 \email{harihars@stanford.edu}
\affiliation{%
  \institution{Stanford University}
  \city{Stanford}
  \country{USA}
}

\author{Jennifer Wang}
 \email{jennifer.wang@microsoft.com}
\affiliation{%
  \institution{Microsoft}
  \city{Redmond}
  \country{USA}
}

\author{Jennifer Wortman Vaughan}
 \email{jenn@microsoft.com}
\affiliation{%
  \institution{Microsoft Research}
  \city{New York}
  \country{USA}
}

\renewcommand{\shortauthors}{}

\begin{abstract}
Despite the widespread use of artificial intelligence (AI), designing user experiences (UX) for AI-powered systems remains challenging. UX designers face hurdles understanding AI technologies, such as pre-trained language models, as design materials. This limits their ability to ideate and make decisions about whether, where, and how to use AI. To address this problem, we bridge the literature on AI design and AI transparency to explore whether and how frameworks \rv{for} transparent model reporting can support design ideation with pre-trained models. By interviewing 23 UX practitioners, we \rv{find that practitioners frequently work with pre-trained models, but lack support for UX-led ideation.} Through a scenario-based design task, we identify common goals that designers seek model understanding for and pinpoint their model transparency information needs. Our study highlights the pivotal role that UX designers can play in Responsible AI and calls for supporting their understanding of AI limitations through model transparency and interrogation.

\end{abstract}





\maketitle

\section{Introduction}

The use of AI technologies has become widespread, from novel systems like machine translators built entirely around a machine learning (ML) model, to advanced features like text auto-completion built into already commonplace applications. Advances in AI are driven by \rv{research and development efforts producing models with various capabilities, but often} disconnected from specific applications or user needs. \rv{For example, AI service platforms~\cite{watson,google-ai,Amazon} such as Microsoft Azure~\cite{Azure} and HuggingFace~\cite{huggingface}  host a growing collection of \textit{pre-trained} models with capabilities in language, vision, audio, and more. There is also a} recent trend of developing large pre-trained models, such as the large language model GPT-3~\cite{brown2020language} \rv{or the multimodal model Dall-E~\cite{ramesh2021zero}}. Fully realizing the potential of \rv{these} new AI technologies requires discovering applications where they can be used to solve user problems and aligning their behavior with user preferences. While these tasks are often an essential part of UX designers' jobs, recent research shows that practitioners grapple with challenges when using ``AI as a design material''~\cite{yang2018investigating,yang2020re,dove2017ux,holmquist2017intelligence}. These challenges discourage the prioritization of UX, leading to failures of AI-driven products and unintended individual and societal consequences. 

Among other challenges, the design ideation process is often hindered by struggles to understand the AI technologies due to their complexities and expertise barriers~\cite{yang2020re,dove2017ux}.  However, effective design ideation does not necessarily require a deep technical understanding of the technology, but rather a ``designerly understanding''~\cite{yang2018machine,yang2018investigating}. What does a designerly understanding of AI involve? Prior work defined it as \rv{the ability to link an} AI technology's capabilities to ways of generating value for users~\cite{yang2018investigating}. Having a good understanding of the design material can enable designers to take UX-led approaches to AI product innovation that prioritize value to users, mitigate potential harms, and better align with the goal of responsible AI (RAI). \rv{However, prior work reported a lack of means to support a good designerly understanding of AI~\cite{yang2020re,dove2017ux}, and to begin with, a lack of knowledge on what designers need for such support.}

\rv{Meanwhile, we recognize that} supporting an understanding of AI has long been the goal of \rv{research on }AI transparency. Besides producing explainable AI (XAI) techniques to illuminate the technical details of models~\cite{gunning2019xai}, the community is moving towards standardized approaches to transparent reporting with \textit{AI documentation frameworks} such as model cards~\cite{mitchell2019model}, datasheets~\cite{gebru2021datasheets}, and AI service factsheets~\cite{arnold2019factsheets}. These frameworks \rv{are often motivated from the perspective of RAI---}to help both practitioners \rv{and end users evaluate the suitability of the model or dataset for their products or contexts.}  This suitability assessment must be supported through \rv{means of} understanding caveats such as unintended use cases, limitations, and potential pitfalls, \rv{in addition to basic details such as model inputs and outputs.}

 \rv{With the rise of pre-trained models, which lower the barrier to AI for practitioners, transparent model reporting is all the} more critical. \rv{But while these} services often target engineers, in practice, it is questionable whether these are, or \textit{should} be, the only roles driving suitability assessment and ideation. With the movement to ``democratize AI,'' it is equally important to lower the barrier to ideating on whether, where, and how to use models appropriately and responsibly. By reducing technical investment overhead, the availability of increasingly \rv{powerful} pre-trained models for product development may create both more opportunities and more responsibility for UX designers to drive innovation.

In this work, we set out to explore how to support design ideation around the use of pre-trained models, focusing on enabling a designerly understanding through model transparency. \rv{Specifically, we introduce a hands-on scenario-based design task and } leverage an example of model documentation as a \rv{design} probe~\cite{hutchinson2003technology}. We explore the utility and gaps of \rv{the documentation's} comprehensive categories of information \rv{to pinpoint designers' information needs for understanding a model to perform design ideation. }

\rv{Our study takes two particular stances to inform future work supporting UX designers to work with AI. First, we prioritize RAI practices that proactively mitigate potential harms of AI technologies during their development and explore designers' role in RAI. Therefore, our study protocol emphasizes investigating how designers use critical information such as the model's limitations to engage in responsible design ideation. Second, to enable a designerly understanding of AI,} we draw on the goal-oriented stance in human-centered approaches to studying explainable and transparent AI~\cite{liao2021human,vaughan2021humancentered,suresh2021beyond}, recognizing that understanding is a means to an end~\cite{keil2006explanation,lombrozo2012explanation}, and effective transparency support must be developed according to the end goals. \rv{ Our analysis distills four common goals that designers seek model understanding for, which future work should aim to support. }

\rv{In short, our work makes the following contributions:}

\vspace{-0.2cm}
\begin{itemize}

    \item \rv{\textit{Identifying new challenges in AI UX design practices}: Our interviews reveal that practitioners frequently work with pre-trained models and that new challenges arise when understanding and designing with these models. Echoing findings in prior work, there is a lack of support for UX-led approaches to product ideation, which is especially critical for the responsible use of pre-trained models.}
        
        \item \rv{\textit{Bridging AI design and AI transparency}: We explore using transparent model reporting frameworks to support design ideation with a pre-trained model. While our study provides evidence of their utility, it also reveals significant gaps and calls for moving beyond static documentation to supporting model interrogation. }  
    
    \item \rv{\textit{Identifying four common goals that designers seek out model understanding for and how to support them}: These goals are to engage in divergent-convergent design thinking and eliminate risky design ideas; to create ``conditional designs'' to mitigate AI's varying impact for different user scenarios; to provide AI transparency to end users; and to negotiate and collaborate with their team to advocate for users.  We pinpoint designers' model information needs for each of these goals, and suggest design guidelines to support them. These common goals also highlight the pivotal role that UX designers can play in RAI with an effective understanding of model limitations. }

\end{itemize}

Below we start by reviewing related work that informed our study, then present our methods and findings. We conclude with a discussion of implications for research and practice.

\section{Background and research questions}

\subsection{Challenges of AI as a Design Material}

 Researchers have investigated the challenges for UX practitioners to work with ``AI as a design material,''  \rv{including} the complexity of the material itself~\cite{dove2017ux,yang2018investigating,yang2020re,subramonyam2021towards}. ~\citet{yang2020re} summarize two sources of AI's distinctive design challenges: 1) the uncertainty surrounding its capabilities, with expansive and evolving algorithmic possibilities; and 2) AI's output complexity, stemming from its probabilistic and adaptive nature. \citet{subramonyam2021towards} contend that, because of the complications of developing models, including choosing from different models, AI does not lend itself to the deterministic ``material'' perspective that designers are used to when working with unfamiliar technologies~\cite{giaccardi2015foundations,robles2010texturing,fernaeus2012material}, but instead \rv{has} its material properties \textit{emergent} from envisioned designs. 

Besides making existing UX methodologies (e.g., prototyping and user testing) challenging~\cite{subramonyam2021protoai,yang2019sketching}, these materialistic complexities give rise to pressing challenges in understanding AI~\cite{yang2020re}. \rv{These challenges are }exacerbated by disciplinary barriers and a lack of support for gaining AI literacy~\cite{dove2017ux,liao2020questioning,lu2021impact}. Interestingly, an interview study with experienced AI designers~\cite{yang2018investigating} suggests that design is not necessarily hindered by a lack of technical knowledge, but supported by a ``designerly understanding'' of the technology, often approached through designerly abstractions (e.g., describing its capabilities in relation to user utility) and design exemplars. 

The struggle to understand AI can hinder design ideation, causing designers to fail to recognize ``low-hanging fruits'' to use AI to solve user problems, grapple with envisioning novel uses of AI, or inadvertently attempt uses that exceed technical feasibility~\cite{yang2020re, dove2017ux}. A survey study published in 2017~\cite{dove2017ux} reported that UX designers were rarely involved in the feature planning stage for AI-powered products, but were limited to working on UI designs. In contrast, a recent study \cite{yildirim2022experienced} with experienced enterprise AI designers suggests that they do engage in defining new systems and processes. These engagements require not only understanding the AI's capabilities and conceptualizing how a design idea would add value, but also viability positioning that justifies the use of AI through expected return on investment.

 We recognize that there are also systemic challenges. While the emergent properties of AI call for UX-led approaches to shape technological choices~\cite{yang2020re,subramonyam2022solving,yildirim2022experienced}, individuals may face upstream battles challenging the current software engineering workflows~\cite{subramonyam2022solving,yang2019sketching,windl2022not},
defeating constraints on time, resources, and incentives~\cite{yang2018investigating,kayacik2019identifying,cramer2019confronting}, and overcoming disciplinary and organizational barriers ~\cite{yildirim2022experienced,liao2020questioning}. While our work seeks to empower designers to drive ideation by supporting their understanding of AI, this goal cannot be achieved without also tackling the organizational challenges.

\vspace{-0.2cm}

\subsection{Information Support for AI Design}

A small but growing area of work on supporting AI designers to overcome the above-mentioned challenges has emerged. Prior research produced tooling to support AI prototyping~\cite{subramonyam2021protoai,malsattar2019designing}, new design processes~\cite{liao2021question,kayacik2019identifying,subramonyam2021towards,giaccardi2015foundations,eiband2018bringing}, and boundary objects~\cite{liao2020questioning,subramonyam2021towards,cai2021onboarding} to facilitate collaboration between designers and data scientists. Our work is most directly informed by related work that focuses on providing information and knowledge support for AI design.

Research and industry have produced numerous taxonomies and guidelines to sensitize designers to both AI capabilities~\cite{yildirim2022experienced,liao2020questioning,dellermann2021future} and the AI design space (e.g.,~\cite{amershi2019guidelines,pair,IBM-AI}). \rv{However,} support for designers to seek information about specific models they work with remains under-explored. A small number of tools have been developed to help designers understand \rv{certain} aspects of a model, such as performance metrics~\cite{yu2020keeping,hong2020human}. Others explored approaches to guide designers in envisioning solutions when working with a model. For example, \citet{hong2021planning} developed an NLP playbook to encourage systematic consideration of errors, based on common failures of NLP models. \citet{subramonyam2021towards} suggest the use of ``data probes,'' example data points and their model outputs, to facilitate design thinking and validation. Similarly, to enable exploring GPT-3's promises for interaction design, \citet{lee2022coauthor} created a dataset of instances from writers working with GPT-3. Using examples to support understanding aligns with the ``material'' design perspective, as understanding of materialistic properties can be achieved by experienced affordances~\cite{doordan2003materials,schon1996reflective,holmquist2017intelligence}.

Also under-explored is investigation into the actual information-seeking processes when designing with AI, except for some first-person account of the challenges~\cite{benjamin2021machine,yang2019sketching}. A relevant work is~\citet{subramonyam2022solving}. By interviewing practitioners, the authors investigated how designers and data scientists overcome expertise boundaries by sharing information through low-level details. Engineers share with designers information about the data used to train the model through dataset documentation and other means, and about model behaviors through example outputs, performance dashboards, demos, and explanations such as feature weights, rules, and underlying assumptions. 

\rv{Like most prior research in this area, ~\citet{subramonyam2022solving} focuses on cases where UX practitioners work with the data scientists who develop the model. We instead investigate designers' information needs when working with \textit{pre-trained} models, which can be third-party models or models handed over after completion, where access to the model developers is unavailable or limited. } We also take an ecological position of human-information interaction research \cite{fidel2012human} that people's information needs are best understood by observing tasks being performed. We hence create a scenario-based design task to investigate designers' needs when designing with an unfamiliar \rv{pre-trained} model

\vspace{-0.3cm}
\subsection{AI Transparency}

To support a designerly understanding of AI, we draw from the literature on AI transparency. To facilitate effective and consistent AI transparency, the AI research community has proposed various frameworks for transparent reporting of data~\cite{gebru2021datasheets,holland2020dataset,bender2018data}, models~\cite{mitchell2019model}, and services~\cite{arnold2019factsheets}, often broadly referred to as ``AI documentation.'' These frameworks include standardized categories of information---such as ``performance metrics, intended use cases, and potential pitfalls'' ~\cite{mitchell2019model}---and guidelines to help AI creators transparently communicate the capabilities and limitations of their models or data. Aiming to support ``responsible democratization of AI''~\cite{mitchell2019model}, AI documentation is intended to support evaluating the suitability of a dataset or model for one's use case, and facilitate accountability and governance. These frameworks are increasingly adopted in industry, especially for third-party AI services. For example, model cards \cite{mitchell2019model} have been implemented by Google Cloud~\cite{google-ai} and Hugging Face~\cite{wolf2019huggingface}, and Microsoft introduced ``Transparency Notes'' for its Azure Cognitive Services~\cite{transparency-note}.

While some researchers have explored the needs of practitioners creating AI documentation~\cite{heger2022understanding,hind2020experiences}, empirical studies investigating \rv{its} use is relatively scarce. Through a think-aloud protocol, \citet{boyd2021datasheets} demonstrates that thoughtfully constructed datasheets can help ML engineers understand and make decisions about ethical problems in training data. A recent study by ~\citet{crisan2022interactive} contends that current AI documentation primarily serves people with ML expertise, while non-experts can benefit from interactive interrogation of an expanded form of documentation. Through a user-centered design study, the authors create a prototype of an interactive model card and make design recommendations, including considering information hierarchies and prioritizing critical information to promote productive skepticism.

Another cornerstone of AI transparency is to support understanding of model behaviors through AI explanations, actively studied in the field of explainable AI (XAI) (e.g.,~\cite{gunning2019xai,guidotti2018survey,lipton2018mythos}). XAI techniques typically address user questions such as ``how does the model make decisions?'' or ``why does the model make this particular decision?'' by revealing the features used by the model, how these features are weighted, or the rules that the model follows. Recent studies report that XAI techniques are increasingly used in industry practice as end user-facing features~\cite{liao2020questioning}, by data scientists to debug models~\cite{hohman2019gamut,liao2021human,kaur2020interpreting}, and shared with stakeholders to verify the models~\cite{hong2020human,bhatt2020explainable}. However, it is unclear if designers utilize---or even have the need for---such technical explanations.

\rv{Our study is also motivated by the goal-oriented stance in research that takes human-centered perspectives on explainable AI ~\cite{liao2021human,vaughan2021humancentered,suresh2021beyond}. Rather than focusing on what aspects of the model can be made transparent, this stance prioritizes articulating the goals that people seek out model understanding for, and centers the development and evaluation of explainability methods around these goals. While several taxonomies of common goals of XAI have been proposed~\cite{suresh2021beyond,liao2022connecting,chen2022interpretable}, other works empirically investigated the goals of a specific group of people. For example, by interviewing data scientists and ML engineers, ~\citet{hong2020human} identified the explainability goals of ML practitioners to be model improvement, knowledge discovery, and gaining confidence. }

\rv{We set out to identify designers' \textit{transparency goals} during design ideation to unpack the requirements to support designerly understanding of AI. We introduce} model documentation as a \rv{design} probe~\cite{hutchinson2003technology}. That is, \rv{besides} the ``engineering goal'' of testing the viability of using documentation to support AI design ideation, we are interested in the ``social science goal'' of understanding designers' needs during ideation and the ``design goal'' of inspiring new approaches to supporting model understanding.

\vspace{-0.1cm}

\subsection{Responsible AI (RAI) in Practice}
Our work also aims to contribute to RAI practices by exploring designers' role in responsibly building AI technologies. RAI is concerned with putting theoretical principles of AI ethics into practice, and proactively mitigating individual and societal harms from AI~\cite{shneiderman2021responsible,rakova2021responsible,benjamins2019responsible}. Recent years have seen a growing interest in studying practitioners' practices, challenges, and gaps in dealing with RAI issues, such as fairness~\cite{madaio2020co,madaio2022assessing,holstein2019improving,deng2022exploring,rakova2021responsible}, transparency~\cite{liao2020questioning,hong2020human,bhatt2020explainable}, and accountability~\cite{raji2020closing,brown2019toward}. These challenges are multi-faceted, ranging from individuals' lack of knowledge support and technical means, to socio-organizational barriers such as lacking incentives and enabling internal structures. 

However, UX designers are not always included in these studies of RAI practitioners. A gap seems to exist between many works advocating for collaboration between ML engineers and designers to create good AI UX, and advocating for designers' role in \rv{mitigating potential harms of AI.} \rv{Meanwhile,} recent work recognizes that RAI is fundamentally about serving stakeholders' needs and values~\cite{madaio2020co,delgado2021stakeholder,sloane2020participation}, a position that is central to the deliverable and methodological toolbox of UX practitioners. Studies of enterprise designers also suggest that designers are deeply concerned about RAI issues such as fairness, transparency, safety, privacy, and data use ~\cite{zdanowska2022study,liao2020questioning,yildirim2022experienced}.

To explore designers' role in RAI and how to support such a role, our study emphasizes the need for supporting a designerly understanding of \rv{a model's} limitations, including failures, biases, and potential harms. This emphasis on both capabilities and limitations also aligns with the intent and design of AI documentation~\cite{mitchell2019model,arnold2019factsheets}

\rv{In summary, drawing on these prior works, our study is guided by the following research questions:}
\begin{itemize}
   \item \rv{\textbf{RQ1}: What are UX practitioners' needs and challenges in understanding and working with pre-trained models, particularly to perform responsible ideation? (Sections~\ref{sec:current}--\ref{sec:goals})} 
 \item \rv{\textbf{RQ2}: To what extent are current model documentation frameworks useful for supporting design ideation and what are the gaps? 
 (Sections~\ref{sec:richdesigns}--\ref{sec:goals})}
  \item \rv{\textbf{RQ3}: What are the goals that designers seek model understanding for and how can they be supported? (Section~\ref{sec:goals})}

\end{itemize}{}

\section{Method}
\label{sec:method}
We conducted interviews consisting of a hands-on ideation task \rv{using} a pre-trained model to solve a user problem, and discussions around participants' own experiences performing design ideation for AI. In the sections below, we first describe the ideation task, then the procedure, participants, and analysis of the interviews.

\subsection{Scenario-Based Design Ideation Task and Artifacts Provided}
We aimed to create an ideation task that could be completed in 30 minutes and generate rich discussions. We therefore chose a user problem scenario that is easily accessible, but has a complex solution space, with multiplex user flows. In the scenario we chose, users of an online microblogging platform share online articles without understanding them or helping their followers understand them, leading to the spread of misinformation. We included a Twitter UI (with minor adaptations, such as changing the brand name) in the task introduction to invoke participants' knowledge about microblogging platforms.  The company running the microblogging platform in the scenario had already paid for an AI service which includes a pre-trained \textit{text summarization model}. Participants were asked to act as if they work for the company and try to \textit{come up with a new feature of the microblogging platform that takes advantage of this available model to solve the article misunderstanding problem.} 

We chose to base the model on the extractive text summarization model provided by Microsoft Azure Cognitive Service\footnote{\url{https://azure.microsoft.com/}} because it is a popular AI service with comprehensive documentation.
Two artifacts were provided to help participants to understand the model: a modified version of the \textbf{model documentation} \rv{from the service (a Transparency Note),} and 20 curated \textbf{model input-output examples} (described below). The documentation covers the major components specified in model reporting frameworks like model cards \cite{mitchell2019model}, including a model description, examples of intended uses, warnings against unintended uses, and limitations highlighting impacting factors (i.e., what factors may impair the model's performance). \rv{Images of the documentation artifact used in the study are shown} in Figure~\ref{fig:doc}-Left. \rv{The content is also provided in Table~\ref{content} in the appendix. }

\begin{figure*}
  \centering
  \includegraphics[width=2\columnwidth]{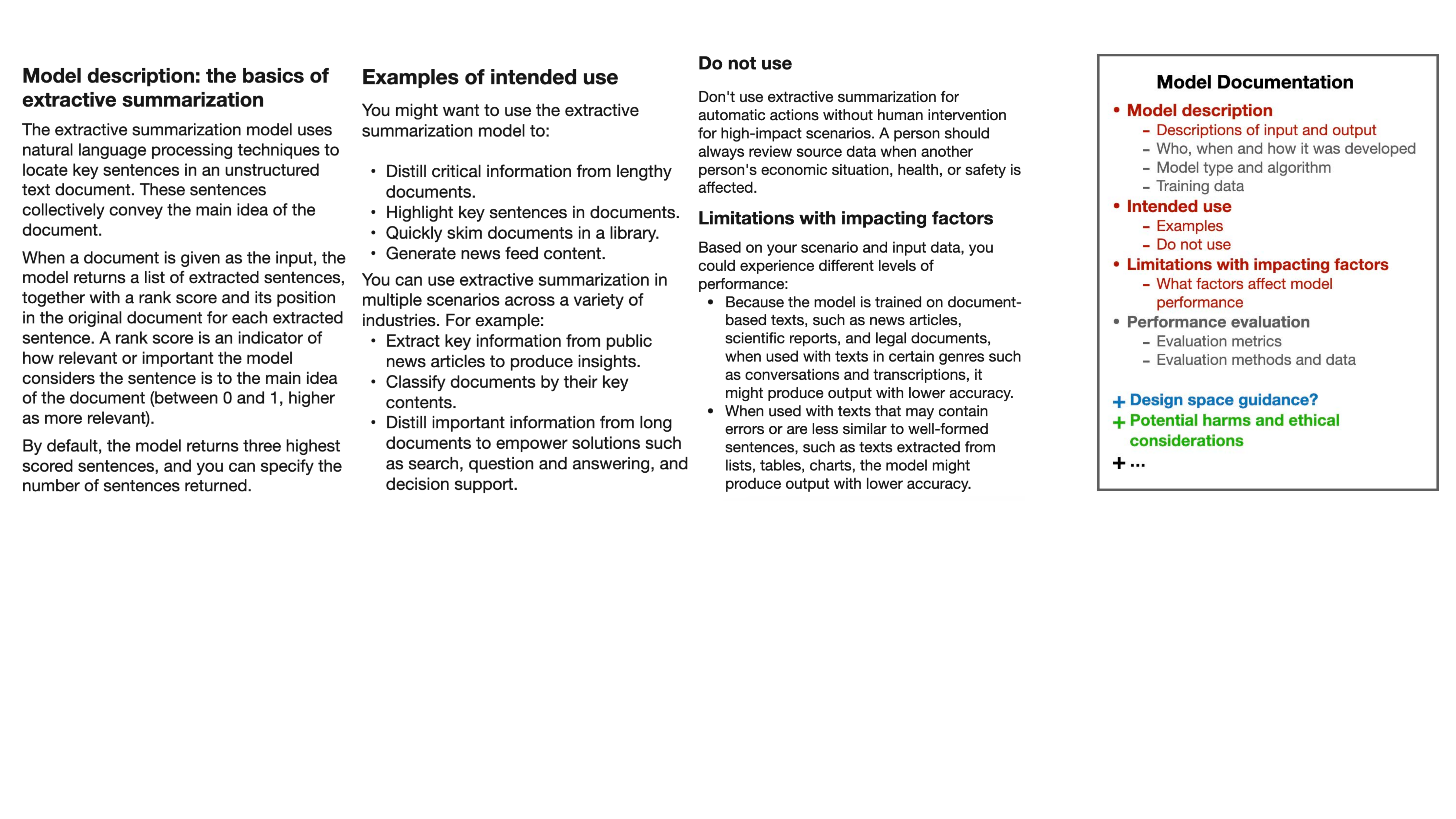}
    \vspace{-11.5em}

  \caption{\textbf{Left}: \rv{Images of the model documentation artifact given to participants, with four categories of model information (content in Table~\ref{content} in the appendix).} \textbf{Right}: \rv{Image of the} summary card shown in the last step for reflection.}~\label{fig:doc}
  \Description{Two images of the artifacts provided to participants. On the left is the model documentation, including model description, examples of intended use, use cases to avoid and limitations with impacting factors. On the right is a summary card of information categories provided and other categories in the model card framework, shown in the last step of the study for reflection}
    \vspace{-2em}
\end{figure*}

The service provides a playground UI \rv{(Figure~\ref{fig:playgrond} in the appendix)} for users to try out the model with their own input examples. Seeing examples of model outputs allows understanding through experienced affordance~\cite{lee2022coauthor,subramonyam2021towards}. For efficiency, instead of asking participants to experiment on their own, we curated 20 online articles from different genres and sources, and of different lengths, and presented them in a spreadsheet with these attributes shown. We then captured their summary outputs from the playground UI, and linked the screenshots to the corresponding input articles in the spreadsheet. The documentation and examples show that the model output includes three components: 
\begin{itemize}
    \item Extracted sentences: Three sentences extracted from the input article that the model identifies as conveying the main topic of the article. 
    \item Rank score: A score indicating how relevant each extracted sentence is to the article's main topic.
    \item Positional information: The position of each extracted sentence in the input article.
\end{itemize}

We chose a summarization model for several reasons. First, language models are among the most popular pre-trained models as they have become increasingly powerful and can be applied to any document input data. Second, to make the ideation task tractable, we opted for a model that has well-scoped capabilities, rather than general-purpose large language models such as GPT-3. We also believe that the relatively narrow scope of extractive summarization was suitable for a task that requires careful ideation to match model capabilities and user needs.

\begin{figure*}
  \centering
  \includegraphics[width=1.55\columnwidth]{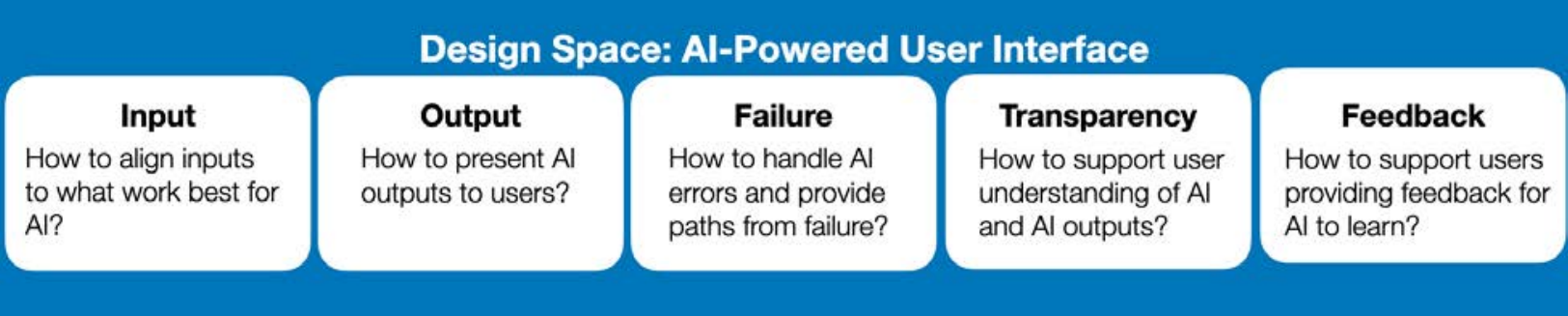}
  \includegraphics[width=1.6\columnwidth]{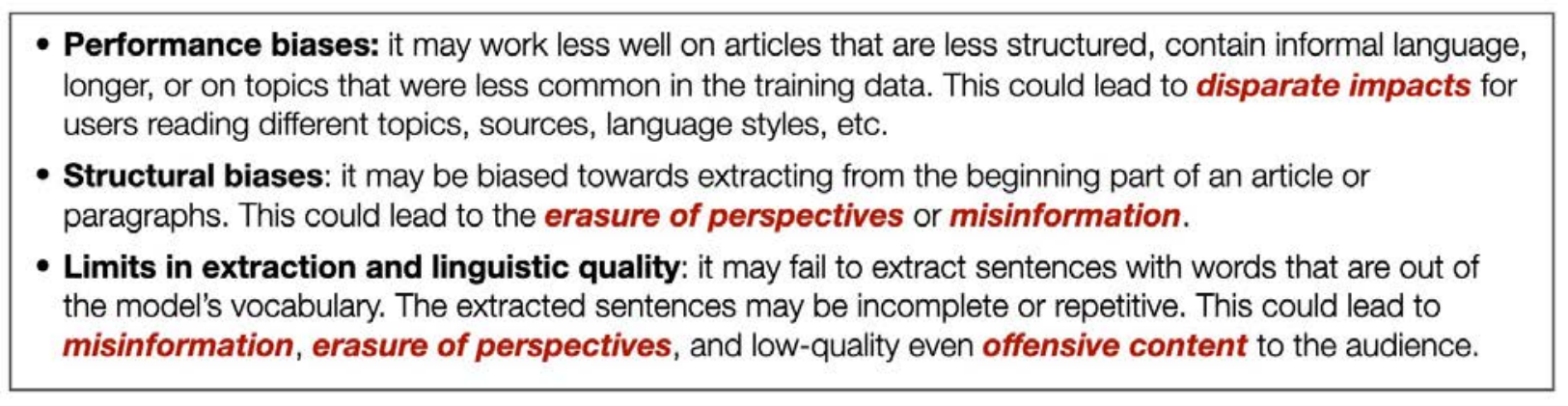}
    \vspace{-0.5em}

  \caption{\textbf{Top}: AI UI design space guidance \rv{provided in the study}. \textbf{Bottom}: Harms considerations \rv{provided in the study.} Each bullet point is a common technical limitation of text summarization models; the potential harms that each limitation can lead to are highlighted in red. \rv{The content is in Table~\ref{content} in the appendix. }}~\label{fig:step3} 
  \Description{Two additional artifacts provided in the study for iteration. On the top is an AI UI design space guidance, suggesting to think through the design of input, output, failure, transparency, and feedback. At the bottom is a set of considerations for the potential harms of summarization models, including disparate impacts, erasure of perspectives, misinformation, and offensive content. }
    \vspace{-1.5em}
\end{figure*}

After participants' initial ideation, we introduced two additional sets of information to help them refine their ideas: \textbf{AI design space guidance} (Figure~\ref{fig:step3}-top), and a list of potential \textbf{harms considerations} (Figure~\ref{fig:step3}-bottom). 

The design space guidance is intended to encourage participants to systematically consider key design elements for AI-powered features, which would allow us to understand their information needs comprehensively. We opted to introduce it after the initial ideation to avoid overwhelming participants. The guidance also serves as additional information about ``what to design''~\cite{buxton2010sketching}. We adapted the ``AI-powered user interface guidance'' in \citet{subramonyam2022solving}, a synthesis of key UI components of AI-powered systems based on 89 industry design guidelines.

The harms considerations were introduced to further investigate participants' ideation around how to use the model responsibly. While transparency on ethical considerations has been a motivating factor for AI documentation frameworks~\cite{mitchell2019model,arnold2019factsheets, gebru2021datasheets}, there is currently no industry standard on how to present them. \rv{We designed the information }based on a review of survey papers mentioning limitations of summarization models~\cite{koh2022empirical,feng2021survey,yadav2022automatic,allahyari2017text} and papers on ethical risks of language models~\cite{blodgett2020language,weidinger2022taxonomy}, as well as discussions with 2 experts of NLP and AI ethics. We chose to lead with common technical limitations of summarization models and highlight the potential harms that each technical limitation can lead to (in red). This delineation between the potential harms and their sources of technical limitations was intended to encourage participants to come up with mitigation strategies that target the sources.

\vspace{-0.1cm}
\subsection{Procedure}
All interviews were conducted online via video conferencing software and lasted around 60 minutes. A {\$50} gift card was provided as an appreciation token for each participant. Participants were asked to read and sign the consent form before they joined the interviews. The study was IRB approved. 

The semi-structured interviews started with a 10-minute discussion of participants' prior experience with designing AI-powered applications. The interviewer probed on how they attempted to understand models in their initial encounters, including their approaches, resources available, and challenges. 

The interviewer then made a 5-minute presentation to introduce the design task described above, including showing the documentation \rv{(Figure~\ref{fig:doc}-Left)} and demonstrating the playground UI \rv{(Figure~\ref{fig:playgrond} in the appendix)}. Participants were then asked to join a FigJam board (whiteboarding feature provided by Figma, a UI prototyping software), where \rv{we provided} the scenario description, documentation, and a link to the spreadsheet of \rv{input-output} examples \rv{obtained from the playground UI}. Participants were encouraged to spend a few minutes to further understand the model by browsing the spreadsheet with \rv{input-output} examples. They were instructed to start ideating whenever they felt ready and follow any processes they usually do. They could use sketching, sticky notes, or any UI widgets on FigJam to communicate their ideas. We also provided a set of microblogging UI components which they could optionally include in their design or annotate directly. Participants were asked to spend no more than 25 minutes on this task, and could stop whenever they were satisfied with their design idea. They were asked to continue thinking aloud throughout the process.

After this, the interviewer asked participants about their perceived understanding of the model, which information they found helpful, and what questions were left unanswered, followed by the two rounds of iteration with the design space guidance and harms considerations \rv{(Figure~\ref{fig:step3})}. For the sake of time, the iterations focused on verbal discussions rather than re-creation of visual designs. Whenever applicable, participants were prompted to reflect on whether the process and information available shared similarities with how they approach AI in their own work

Lastly, the interviewer asked participants to reflect with a summary card as shown in Figure~\ref{fig:doc}-Right. The card listed all the categories of information provided in the task in color, with additional categories that appear in model cards~\cite{mitchell2019model}, the most established model documentation framework, in grey. The latter were described as ``more technical information'' that we excluded (also excluded in the original service documentation). The interviewer asked questions to prompt reflection, such as which category was helpful, whether the grey categories were desired, and what other information they wished to have. The interviewer also introduced the concept of \textit{designerly understanding of AI}---understanding a model well enough to be able to use it as a design material to solve user problems---and asked participants to reflect on what could help them better achieve a designerly understanding in general.

\vspace{-0.2cm}

\subsection{Participants}
\label{sec:participants}

Recruitment was carried out through two routes. First, recruiting messages were disseminated in a large international technology company's UX-focused online communities, across product lines and locations. Second, the authors posted recruiting messages on Twitter and LinkedIn. The messages called for participation of people who are in roles that perform design ideation often (including designers, UX researchers, and product managers (PMs)), and have experience working with AI. We limited to these groups since we are interested in learning about participants' own experience ideating for AI-powered products. 

The interview study included 23 participants (8 male, 15 female), with 17 recruited via the first route and 6 via the second. Participants from the same large company were distributed in 6 countries with no overlap of first-line teams. The remaining participants work in a mix of large companies, start-ups, and non-profit organizations. 

When participants signed up, they were asked to fill out a form that gathered information about their \rv{demographics}, professions, and their self-reported experience with designing AI and NLP powered applications, respectively (never / limited experience / part of my day-to-day job / I consider myself an expert). The majority of participants have designer titles (N=17), while 3 are HCI or UX researchers, and the remaining 3 are PMs. Detailed information about the participants can be found in Table~\ref{demographics} in the appendix. The last two questions were used to group participants into more or less experienced groups with AI design. Overall, we considered 6 participants to be in the less experienced group; these participants either answered ``less experienced'' or ``never'' to both questions regarding AI and NLP or confirmed in the interview they never designed AI-powered features in their job. 
%\hari{Should we just insert the participants' table here instead of the appendix?}
% 11 of them had 1--5 years of experience in the profession, 8 had 5--10 years, and the remaining 4 had more than 10 years.
\vspace{-0.2cm}
\subsection{Analysis}
\label{analysis}
Interview transcriptions included question-answering and think-aloud data. Coding started with the first and second authors performing open and axial coding informed by Grounded Theory research~\cite{corbin2015basics} on a common set of 5 interviews. They discussed and converged on a set of axial codes, with which the first author continued coding the rest of the interviews. The axial codes will be highlighted in \textbf{bold} when discussing findings. After that, the first author performed a first round of selective coding to identify themes, then \rv{iteratively} presented to the other authors for feedback.

Through a human-information interaction lens~\cite{fidel2012human}, we paid particular attention to places where participants showed their attention to, perception of, use of, and feedback for the categories of \rv{model} information provided. These include what they commented on while reading the documentation, what appeared in their think-aloud comments while performing the task, and their answers to the reflection questions after the task. We coded both the \textit{categories of information} and participants' \textit{goals} behind the information sought. We also mapped the relations between the two with the axial codes. These results \rv{are} presented in Sections~\ref{sec:goals}.

\section{Findings}

 \rv{Since the focus of our study is on design ideation with pre-trained models, we first situate our results by discussing what this task currently looks like in practice (RQ1). We then present a brief overview of participants' designs, demonstrating that they were able to engage in design ideation supported by the documentation (RQ2), but their model understanding and design outcomes varied by their level of experience with AI design (RQ1). Finally, in Section~\ref{sec:goals}, we present our main results, identifying four common goals that designers seek model understanding for, pinpointing designers' model information needs for each goal including gaps in the current documentation, and suggesting design implications to support them (RQ1, RQ2, RQ3). Throughout the findings, we highlight the pivotal role that UX designers can play in RAI with an understanding of model limitations.}

\vspace{-0.1cm}
\subsection{\rv{Putting Design Ideation with Pre-Trained Models in Context:} Current Practices \rv{(RQ1)}}
\label{sec:current}

\rv{
As described below, while previous HCI research focused on supporting UX practitioners to work with data scientists, we found that, on the ground, practitioners also frequently work with pre-trained models without the direct involvement of the data scientists who built them---a trend that may increase as more powerful pre-trained models such as GPT become widely available. 
This makes it more challenging to assess models for suitability and ideate on how to use the models to solve user problems, since information cannot be obtained directly from the data scientists involved. Unfortunately, at the current time, designers often do not play a central role in the ideation stage for AI-powered products and do not have the information support to obtain a good enough understanding of models to engage in effective ideation.
}

\subsubsection{How is design ideation currently performed in practice?}

Our study presented participants with a scenario that requires figuring out how to use a given model to solve an existing user problem. We found that \rv{participants} (or their teams) commonly face this type of scenario \rv{in their day-to-day practices}. \rv{Namely,} \textbf{practitioners frequently work with third-party models} \rv{(N=12)}---sometimes referred to as ``out-of-box'' or ``off-the-shelf'' models---to add AI-powered features to existing \rv{products}.  

\rv{Especially in larger companies, practitioners also strive to \textbf{reuse AI capabilities} that the company already owns, whether purchased from third parties or developed by R\&D teams.}
Echoing previous studies, there is still a common separation of design and model development~\cite{subramonyam2022solving}. Designers do not necessarily distinguish between working with a pre-trained model or an ``in-house'' model handed over \textit{after} its completion, as in several cases \rv{(N=5)}, participants could not recall the sources of the models they worked with. \looseness=-1

Curiously, 9 participants mentioned they or their teams engaged in various degrees of \textbf{exploration around the use of recent large pre-trained models}, such as GPT-3 and Dall-E, including attempts to define product features and tinkering with the APIs \rv{using playground UIs} on their own. However, when probed further, none could clearly articulate a ready outcome or an established process to explore these models, showing that ideation on how to use large pre-trained models is an emerging task that raises much interest but \rv{is} still challenging.

Similar to previous studies~\cite{dove2017ux}, we found that UI/UX designers are often not the drivers for product or feature definition, though they are more likely to be in smaller organizations or start-ups (P12, P15, P16)~\cite{yildirim2022experienced}. In large companies, \rv{this} task is often led by PMs, with input from designers. 
%P2, a UX researcher, mentioned that their research is often dependent on a PM's choice of models: ``\textit{There were many different off-the-shelf models. And so we asked, which one would they like to focus on and that decision was purely made by them.}''
Participants frequently expressed \textbf{ dissatisfaction in being excluded in the ideation stage}, as ``\textit{I think we should be because we're gonna carry all that implications of a technology choice}'' (P17). Participants  attributed their lack of means and motivation to understand models to this separation between UI design and ideation, \rv{as} ``\textit{they show up and someone already said this is the problem and this is the solution and you feel like you haven't had a stake, or haven't had a chance to research that problem for your own understanding}'' \rv{(P7).} Participants also described their experience of \textbf{design failures due to this siloed process and a lack of model understanding}, as: ``\textit{I came in a later stage. The PM had already defined all the specs. I mapped out the ideal customer journey and a service blueprint...it turns out we don't have the technical feasibility to cover all of them}'' \rv{(P8).}

%We also believe that a better understanding of available models could better equip designers to drive feature ideation tasks. Designers (P3, P9) who are more engaged in these tasks are among 

\vspace{-0.1cm}

\subsubsection{What are the current approaches to obtaining a designerly understanding?}
 About half of participants have done so by \textbf{reading some form of model documentation}. While a few mentioned formal documentation of third-party AI services or GPT-3, designers often rely on notes written by PMs or model developers. Participants expressed \textbf{struggles with digesting documentation}, because ``\textit{they are explaining very complex things and most of them are just plain text}'' (P14). Some mentioned that \rv{creating high-quality }documentation is usually not \rv{a priority} due to resource constraints.

Participants also sought ``experienced affordance'' by \textbf{examining model inputs and outputs}. P3, P9, and P20 mentioned ``playing around'' with GPT-3 through the playground UI. However, \textbf{means to directly interact with models are often unavailable}. Instead, participants mentioned that their initial encounters with models involved a demo from data scientists or third-party sellers showing examples of inputs and outputs, or being given examples together with documentation. This lack of direct access is common even for in-house models. P10 and P21 approached this challenge by curating their own ``\textit{golden set of inputs}'' and obtaining outputs from the engineers to support their understanding and design decisions. Several participants \rv{(N=5)} expressed excitement upon seeing the playground UI in our design task, showing a gap in their current practices and a strong desire to tinker with models directly. 

Lastly, \rv{as the majority of participants have also worked with ``in-house'' models built by data scientists, in these cases, they} \textbf{learn about models by talking to data scientists}. When performing our design task, they frequently described the experience of reading documentation unfavorably compared with that of speaking to a data scientist directly.

\vspace{-0.2cm}
\subsection{ \rv{Overview of the Design Outcomes (RQ1\&2)}}
\label{sec:richdesigns}

To \rv{investigate the feasibility of supporting design ideation with the documentation framework (RQ2), and to} ground our later discussions of participants' information needs and goals, we briefly overview the variety of design ideas that participants came up with. \rv{To shed light on the challenges (RQ1), we also} highlight differences between designers with more or less experience with AI.
\vspace{-0.1cm}
\subsubsection{Participants created rich designs with \rv{various details}.} \rv{With the same set of provided model transparency artifacts, participants arrived at different designs to address the scenario. }17 out of 23 participants presented a feature that shows AI-generated summaries together with shared articles. 7 participants explored a feature that \rv{nudges} the user to understands the article content before sharing. 4 participants discussed a feature that uses an AI-generated summary to help users to write their own summary.

\begin{figure*}
  \centering
  \includegraphics[width=2\columnwidth]{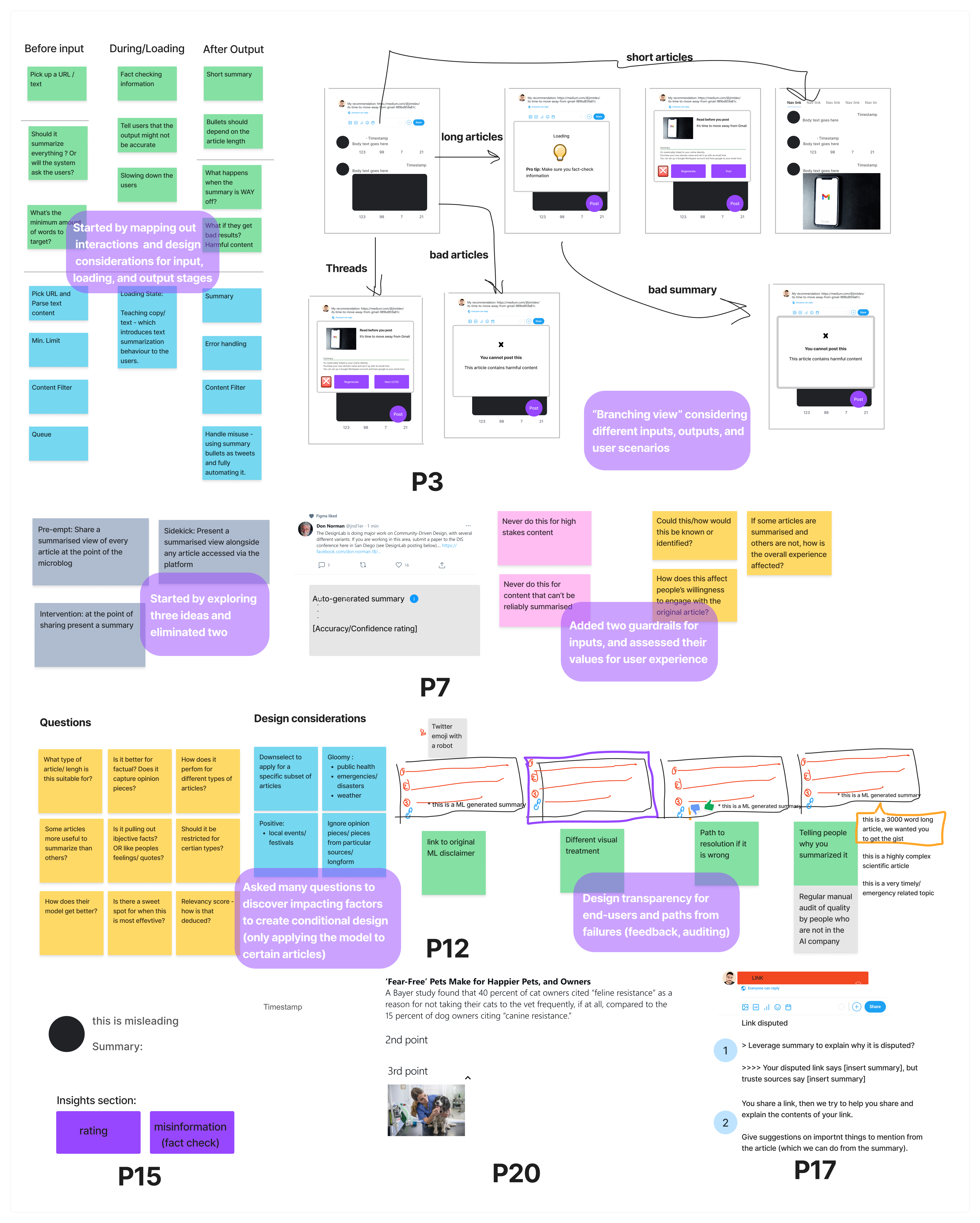}
    \vspace{-1em}

  \caption{Example designs created by participants. The bottom three are from participants in the "less experienced with AI design" group.}~\label{fig:designs}
  \Description{Six example designs from participants. The top 3 are from more AI-experienced designers (P3, P7 and P12), and the bottom 3 are from the less experienced designers (P15, P20, P17). Their contents are explained in Sections 4.2 and 4.3. }
    \vspace{-1em}
\end{figure*}

Participants' designs \rv{also had rich details}. For example, P3, P7, and P12, who are among the most experienced AI designers we interviewed, created sophisticated designs (Figure~\ref{fig:designs}). P3's designs considered different conditions: applying the AI to only longer articles; quality-checking the original articles and summaries to prevent disputable content from being shared; and a pop-up window to view summaries in sequence for users who share multiple articles in a thread. P7 added \rv{user-facing} transparency elements about the model's accuracy, confidence, and explanations, and added that their feature should not be applied to high-stakes topics. P12 presented a summary feature that should only be applied to factual articles but not opinion pieces. Several layers of detail were added: a link to the original article, a disclaimer indicating this content is AI-generated, paths from AI failures including feedback and model auditing, and an explanation for why a summary is provided. 

%\hari{I wonder if it's worth categorizing the details here and connecting them to experience. There is something to be said about experiential knowledge in interpreting facts from model documentation and filling in the gaps which novice designers may not be able to do well. Would it make sense to switch the content within the two sub-headings? I think we can better synthesize details-experience connections.}

\vspace{-0.2cm}

\subsubsection{\rv{Less experienced AI designers faced more challenges approaching model }understanding and ideation.}

As described in Section~\ref{sec:participants}, we identified a sub-group of 6 participants (P8, P11, P15, P17, P18, P20) who are less experienced with AI design. We compare their designs to the more experienced AI designers'. We note that although this group had limited experience with designing AI features in their jobs, they still showed a level of knowledge of and strong interest in AI. Our comparison is \rv{not intended }to generalize the relationship between AI experience and ideation outcomes.          

We observe two clusters of designs created by the designers less experienced with AI. One cluster (P15 and P20, whose designs are in Figure~\ref{fig:designs}, and also P8) presented relatively simplistic designs, without as many details as in the designs by the experienced designers. Interestingly, all of them opted to pick one single example out of the playground UI outputs, and created visuals around the content. The other cluster (P17 as in Figure~\ref{fig:designs}, and also P11) had a distinct pattern of quickly generating multiple ideas, with some diverting from the common ideas participants converged to, but under-exploring the feasibility of these ideas with the given model. For example, as shown in Figure~\ref{fig:designs}, P17 suggested a feature that would leverage a summary to identify and explain links with disputable content. 

Furthermore, we observed that the less experienced group was significantly more likely to skip the step of examining multiple examples from the playground UI---83.3\% versus 11.8\% among the rest of the participants. \rv{Despite the evidence of less effective understanding and ideation, they} were more likely to say yes when asked whether they felt they had a good understanding of the model---50\% versus 11.8\% among the rest of the participants. 

\rv{In short}, the more experienced participants sought information more thoroughly. \rv{With a better understanding of the model, they were able to generate more} sophisticated and complete design ideas that are grounded in technical feasibility. However, they also tended to find the provided information inadequate to support what they intended to design. \rv{In the next section, we unpack what additional information is required and why.}

\vspace{-0.1cm}

\subsection{Common \rv{Transparency} Goals \rv{and Information Needs (RQ1, RQ2, RQ3)}}
\label{sec:goals}

\rv{In this section, we present our main findings. To answer RQ3, we identify four common goals for which designers sought out model understanding. These goals are not meant to be mutually exclusive. To answer RQ2, for each goal, we pinpoint which categories of model information in the provided documentation were used, and what missing information was requested. We also suggest design implications to support each goal to answer RQ1. These results are summarized in Table~\ref{summary}. In the appendix, we provide more details on how each category of model information was used or sought, divided in two tables---Table~\ref{tab:information} for provided information and Table~\ref{tab:information-2} for additional information requested.}

\begin{table*}[]
\sffamily
    \centering
    \begin{tabular}{cccc}
        \rv{Transparency goal} & \rv{Provided info used} & \rv{Requested information} &  \rv{Design implications to support the goal}  \\
          \toprule
         \begin{tabular}{>{\centering\arraybackslash}p{2cm}}\rv{G1: Divergent-convergent design thinking} \end{tabular} & 
         \begin{tabular}{>{\centering\arraybackslash}p{3cm}} \rv{intended uses, model description, input-output examples, harms considerations, unintended uses, limitations}  \end{tabular}& \begin{tabular}{>{\centering\arraybackslash}p{3cm}}\rv{output analysis, explanations}\end{tabular} &  
         \begin{tabular}{>{\arraybackslash}p{6.6cm}}
        \rv{ - Inspire divergent design thinking with example use cases and user contexts, such as providing or helping define user workflows or scenarios.}
         
        \rv{ - Scaffold defining UX risks by supporting discovering a broad range of model limitations and providing ethical consideration guidance.}
         
        \rv{ - Support risk-benefit assessment of candidate design ideas; develop risk-oriented evaluation metrics and practices.}
         \end{tabular}
    \\
      \midrule
      
        \begin{tabular}{>{\centering\arraybackslash}p{2cm}}\rv{G2: Conditional design} \end{tabular} & 
         \begin{tabular}{>{\centering\arraybackslash}p{3cm}}\rv{impacting factors in limitations and harms considerations, examples of different categories} \end{tabular}& \begin{tabular}{>{\centering\arraybackslash}p{3cm}}\rv{training data, explanations, disaggregated evaluation with performance and other output characteristics, confidence/ uncertainty} \end{tabular} &  
         \begin{tabular}{>{\arraybackslash}p{6.6cm}}
         \rv{- Facilitate discovering and testing hypotheses about impacting factors and assessing their UX impact.}
         
         \rv{- Support decisions about whether to create conditional designs by  exploring the design space (e.g., creating intermediate prototypes) and assessing their user values.}
         
         \rv{- Develop design patterns and implementation guidance for conditional designs.}
         \end{tabular}
    \\
     \midrule
        \begin{tabular}{>{\centering\arraybackslash}p{2cm}}\rv{G3: Transparency for users} \end{tabular} & 
         \begin{tabular}{>{\centering\arraybackslash}p{3cm}}\rv{ model description, limitations, harms considerations} \end{tabular}& \begin{tabular}{>{\centering\arraybackslash}p{3cm}}\rv{performance, confidence/uncertainty, explanations} \end{tabular} &  
         \begin{tabular}{>{\arraybackslash}p{6.6cm}}
         
        \rv{ - Allow seeking model information by user questions or needs, such as re-structuring the documentation and providing other information channels.}
        
        \rv{- Support translating information in documentation to transparency designs for users.}
         \end{tabular}
    \\
      \midrule

            \begin{tabular}{>{\centering\arraybackslash}p{2cm}}\rv{G4: Team negotiation and collaboration} \end{tabular} & 
         \begin{tabular}{>{\centering\arraybackslash}p{3cm}} \rv{harms considerations, limitations, design space guidance} \end{tabular}& \begin{tabular}{>{\centering\arraybackslash}p{3cm}}\rv{customizability and improvability, algorithm, training data and other development information} \end{tabular} &  
         \begin{tabular}{>{\arraybackslash}p{6.6cm}}
         
     \rv{- Empower designers by prioritizing their suitability assessment of the model for the users, with both informational and organizational support.}
     
    \rv{ - Equip designers to collaborate with engineers with technical literacy, actionable suggestions for model improvement, common references, and boundary objects.}
         \end{tabular}
    \\
      \midrule     
         
    \end{tabular}
    \caption{\rv{Summary of goals for which designers sought out model information, what information they sought in the provided model transparency artifacts and what is missing, and design implications for supporting each goal}}
    \vspace{-0.5cm}
    \label{summary}
\end{table*}{}

\subsubsection{\rv{Goal 1 (G1):} To engage in divergent-convergent design thinking \rv{and eliminate risky design ideas}} We frequently observed the diamond process~\cite{diamond} of design thinking \rv{(N=14)}, with a divergent stage of generating many potential design solutions followed by a convergent stage of refining them. \rv{As described below, this process drove designers to seek model understanding necessary for assessing potential UX benefits and risks that would arise from their designs.}

\rv{For the divergent stage, participants often (N=6) found the section with examples of \textit{intended uses} helpful to ``\textit{jump-start}'' (P9) generating design ideas. We also observed a common strategy (N=5) of \textbf{delineating user workflows} as a way of ideating on potential places for a summarization feature.  }
 
 \rv{For the convergent stage, a} \textbf{risk-benefit analysis} was often performed to eliminate solutions that are risky or deliver less value to users.  This process is best illustrated in P7's design shown in Figure~\ref{fig:designs}. P7 started by generating three possible solutions that they called preempt, intervention, and sidekick. Their convergent process required understanding how reliable the model is, which they \rv{approached} by examining playground examples: ``\textit{I have the question of how reliably it could perform [for different designs]... if it was an intervention and it was unreliable...you're out of your extra step and it's literal nonsense. And that really diminishes somebody's experience with the whole product, so that presents, I think, a huge risk.}'' Later, as they proceeded with the sidekick idea, they realized that the benefit provided might be limited and they re-visited the other ideas:  ``\textit{now that I'm fleshing this out, it's making me feel this would make people read things even less, because just anecdotally for myself, if I saw this I would definitely not click the article.}''
 
\rv{Performing the risk-benefit analysis led participants to seek an understanding of model capabilities based on \textit{model descriptions} and \textit{input-output examples}, and model limitations from  \textit{harms considerations}, \textit{impacting factors}, and \textit{unintended uses}.} \rv{However, translating model capabilities into UX benefits, and model limitations into UX risks is a non-trivial task.}
\rv{One common translation strategy observed in the majority of participants is to examine the examples and \textbf{mentally simulate how users would perceive and react} \rv{to them}.  As discussed in Section~\ref{sec:richdesigns}, the experienced AI designers often examined  \rv{a mix of} input examples that are representative of articles shared on the platform, examples from multiple categories, \rv{and} edge cases in the hope of revealing model limitations. However, not all participants engaged in these productive strategies.}  

\rv{When translating model limitations to UX risks, participants often (N=6) expressed confusion about \textbf{what failures meant from a UX perspective}. Failures can arise not only due to poor performance as measured by standard performance metrics---the focus of the limitations section in the provided documentation---but can also be caused by other characteristics of the model outputs (like being too long or not coherent enough for a design) or API properties (such as speed or reliance). As a result, participants found the current limitations section inadequate and engaged in \textbf{discovering additional model limitations} that can cause UX failures.}

\rv{Participants appreciated the \textit{harms considerations} to help them think through potential negative consequences. Remarkably, many participants (N=7) took it as inspiration to \textbf{anticipate harms specific to their design and users}. Upon reading about performance biases, for example, P7 questioned the downstream harm of their own design: ``\textit{Am I creating a potential skew? All articles that are extremely neat get a good summary and any that are too complicated have a low accuracy. It's possible...all the ones that are summarized well are click-baity articles...does that really help?}'' } 

\rv{To anticipate UX benefits and risks,} some participants \rv{also requested \textit{descriptive analysis of outputs} to understand the general characteristics of model outputs} such as the distribution of output lengths and \rv{frequencies of certain types of words.} \rv{Such information} should ideally be provided with regard to \rv{input articles specific to} their product. In addition, some participants sought \textit{model explanations} about what features the model relies on. For example, P1 asked ``\textit{does it give more importance to numbers?}'' \rv{as they reasoned that numbers may then show up more often in the summaries.} \rv{Such requests were frequently triggered by observing distinct or unexpected model behaviors in input-output examples.}

\rv{Finally, participants expressed a desire to more \textbf{accurately assess the candidate design ideas by risk}, such as through user testing.} P7 called out a need for risk-oriented evaluation metrics rather than traditional UX metrics: ``\textit{I would want to create different concepts and evaluate them through these lenses [of risks]... typical design world you would say, oh, is this good or not, do users like it or not. But I think this would be the other test. So looking at the potential risks in doing something a particular way}.''

Design implications based on these findings are summarized in Table~\ref{summary}.
\vspace{-0.2cm}

\subsubsection{\rv{Goal 2 (G2): }To create ``conditional designs'' \rv{to mitigate AI's varying impact for different user scenarios}}
\label{sec:conditional} We observed that participants frequently \rv{(N=9)} approached \rv{the model's} output uncertainty~\cite{yang2020re} by creating \textbf{different designs for different types of inputs} or \textbf{different types of outputs}. \rv{They often did so by putting guardrails on inputs and outputs---only applying the model to inputs that it is reliable for, or blocking problematic outputs. Other common conditional designs included triggering user warnings in certain conditions or providing user controls such as ``\textit{toggle on or off this feature for those articles}'' (P12).} These observations suggest that experienced AI designers are mindful about moving away from only focusing on golden paths or ideal hero scenarios---a ``traditional'' design practice that may lead to failure of AI UX~\cite{hong2021planning}. This type of design thinking also stems from a common practice of \textbf{designing for different user scenarios}. We refer to these common processes of creating different designs for different conditions as \textit{conditional designs}. Such designs are best illustrated in the example of P3 (Figure~\ref{fig:designs}) who considered different designs based on article length (input) and summary quality (output), as well as for users who share many articles in a thread (scenario).

This goal of creating conditional designs gave rise to participants' pronounced needs to understand impacting factors, \rv{which appeared in the \textit{limitations} and \textit{harms considerations} sections}. There are several challenges. First, not all impacting factors are available in the documentation. Participants often discovered new factors \rv{of interest} by considering different user scenarios. Participants therefore expressed desires to \textbf{discover or verify more impacting factors}. \rv{While many participants (N=8) attempted this by visually examining \textit{\rv{input-output} examples} of different categories, this approach might not lead to an accurate understanding, but ``\textit{create a cognitive load}'' (P2). Some participants requested information about} \textit{training data}, \textit{explanations}, and \textit{disaggregated evaluations}~\cite{barocas2021designing} (also referred to as sub-group analysis) \rv{to help them discover impacting factors}. As illustrated in P12's examples in Figure~\ref{fig:designs}, they started by asking many questions regarding unknown impacting factors, such as ``\textit{what type of article length is this suitable for?}'' and ``\textit{is it better for factual?}'',  then probed on the explainability-related ``how'' question to infer potential factors: ``\textit{is the summary only pulling out objective facts? Or also peoples' quotes?}'', and questions suggesting a desire for disaggregated evaluation: ``\textit{how does it perform for different types of articles?''} At the end, P12 reflected that they did not have a good understanding of the model, and made an assumption that they should avoid summarizing opinion pieces based on observations of example outputs.

\rv{The decision to create a conditional design must be carefully justified. It comes not only with a development cost but also a cost to UX, since it can create ``\textit{an odd feeling and inconsistency}'' (P7). Participants wanted to \textbf{assess the impact} of factors of interest---both on } performance metrics \rv{and on} other output characteristics \rv{like} the structural patterns and content quality. \rv{This justification must also be assessed with regard to the actual \textbf{user benefits of a conditional design}, considering factors such as  frequency and user contexts of a given condition.} For example, P7 opted to ignore \rv{the concern about} low-quality \rv{summaries for articles about traveling, based on observing an example}, since the consequence of misinformation about such articles might be less serious. After reading about the impact of \rv{lack of article structure on performance} in the limitation section, P9 decided to ignore it, \rv{saying} ``\textit{if articles that are being shared on the microblogging site were mostly chart heavy, like scientific publications, then I would have more questions.}''

Finally, \rv{designers struggled with \textbf{how to implement conditional designs}, often} realizing that another technical component might be required, like a separate model to detect the article genre.  Participants commonly \rv{(N=6)} sought to create guardrails on outputs by leveraging the model \textit{confidence or uncertainty}, prompting them to look into whether the model can provide that information.

Once again, design implications derived from these findings are summarized in Table~\ref{summary}.

\subsubsection{\rv{Goal 3 (G3):} To provide AI transparency to end users} \rv{To create ``\textit{interaction-level interventions}'' (P9) to mitigate the harms of designs, another common goal is to \textbf{transparently expose limitations
and potential harms to end users}.} \rv{Many participants (N=10) attempted to meet this goal by \textbf{incorporating information from the documentation into their designs}, including information from the \textit{limitations} and \textit{harms considerations} sections. Some further requested to expose information about \textit{performance}, output \textit{confidence or uncertainty}, and \textit{explanations} of how the text summarization works. However, participants faced challenges in \textbf{how to translate and effectively present this information in the UI}:}
 ``\textit{we definitely need to surface this information in the documentation. But the key question is from a UI perspective, what needs to change, right?... you can have a pop-up appearing over there highlighting that this capability is in preview, and there could be certain limitations... Click here to learn more and then you get to the documentation}'' (P10).  

Some participants also discussed that in their own work, to be able to effectively communicate the model information to end users, they need to seek an understanding of the model by \textbf{asking questions on behalf of the users} to data scientists\rv{, which they are unable to do with documentation alone}: ``\textit{I think a lot of the questions about the human impact of the model are very much within the designers' purview to ask questions to 
the ML team...the hard part is that a lot of them are not visible. It's such an intangible thing... you have to be really familiar with the material to be able to even have coherent thoughts about it}'' (P12). In other words, an AI-powered product cannot be truly transparent and supportive of user understanding if the designer themself lacks an understanding of the AI material they are working with.

\subsubsection{\rv{Goal 4 (G4):} To negotiate and collaborate with their team \rv{to advocate for users}} \rv{When probed about designers' role in creating responsible AI products, besides design interventions, participants emphasized designers' responsibilities in \textbf{advocating for users} by anticipating potential harms to users and communicating them to their teams, including \textbf{pushing back on the use of a technology}. These points are best made by P9: ``\textit{I think designers are probably in the best position to explain those problems back up the chain because we have good tools about modeling users and contexts in scenarios. So we can say, hey, have you thought about the single mom who's looking at this interface and how it presumes she has a husband and how offensive that is? That ability to frame that as a human problem as opposed to a business problem might have more of an influence within the conversations of an organization. And I think that we end up being a small amount of gate keeping for the vetting of software. So if we found that something was actively propagating misinformation we can reject it... we have to advocate for users, not just business outcomes.''} To engage in such advocacy, designers can ``\textit{feel empowered}'' (P23) by having a good understanding of model limitations. Indeed, we often observed participants pushing back on the use of the model after reading the \textit{harms considerations}.}

\rv{Designers also actively seek to \textbf{collaborate with engineers to resolve technical limitations and improve the model} for their product. This tendency prompted several participants (N=5) to request additional information about model \textit{customizability and improvability}, such as whether it is possible to gather domain-specific data to fine-tune the model. The service roadmap information about future updates is also important for coordinating with the team so they know whether ``\textit{some constraints [will be] erased [in the future]}'' (P13). P2 and P6 also appreciated the format of the provided \textit{harms considerations}, in which different sources of technical limitations were delineated, so that they could work with the team to ``\textit{address each of these different sources of possible harms}'' (P2).
 }

\rv{Lastly, we highlight two additional roles that documentation can serve to facilitate team coordination. One is that a documentation} can be used as \textbf{common references and boundary objects} to support collaboration, \rv{especially the sections on \textit{harms considerations} and \textit{design space guidance}}, as illustrated in P7's response: ``\textit{I'm wondering how much of this is just generally useful for a team. Obviously the model stuff is true, but some of this guidance around human-AI interaction could also be useful for everybody to be mindful of.}'' Second, participants deemed documentation a useful resource to help them to \textbf{develop general AI literacy} \rv{for} effective cross-team communication in the long run, which motivated them to seek more technical information in the documentation such as the \textit{algorithms} \rv{used}, \rv{information about the} \textit{training data}, \rv{and other details about \textit{model development}}: ``\textit{it's also about educating the designer [about] different types of learning models and algorithms, so that when we communicate with data scientists, we can use the same language and talk about the same thing...  on the long term I feel [documentation] also should be about education}'' (P5).

\section{Discussion}

\rv{We have identified UX designers' diverse information needs when working with pre-trained models as design material, and how these needs are engendered by their task-specific, role-specific, and socio-organizational goals. The results demonstrate the utility of model documentation in sensitizing designers to the capabilities and limitations of a model for design ideation, but also reveal many gaps. We found that designers gravitate towards critical information that helps them understand model limitations and adopt a set of strategies to mitigate the negative user impact of these limitations. In this section, we discuss future directions for supporting ideation with model transparency and argue for better engaging designers in RAI practices.}

\subsection{\rv{From Model Documentation to Model Interrogation}}

 While there has been extensive work on AI documentation~\cite{gebru2021datasheets,mitchell2019model,arnold2019factsheets,hind2020experiences,heger2022understanding,aboutml}, who the consumers are and how they are consuming it has not been well studied empirically. Our work serves as a case study to explore the model reporting needs of UX practitioners. We found that they can benefit from AI documentation, and are already consuming it on the ground.  \rv{However, only 5 out of 23 participants answered affirmatively that they understood the model in our study well. Participants requested additional categories of information, and some struggled with a lack of complete or concrete understanding of provided information. }

\rv{ Their struggles were not due to a lack of ML expertise \cite{crisan2022interactive}, as participants had little difficulty comprehending the documentation (though this may not generalize to designers with little knowledge about AI). Instead, the challenges arose when \textit{contextualizing} model information for their setting and users.  It is impossible for documentation creators to anticipate every downstream use case. This suggests that we should provide opportunities for designers to interrogate the model with their own data instances, factors of interest, hypotheses, and questions. Additionally, static documentation falls short in supporting the \textit{co-evolving of design solutions and model understanding, } which can be seen as an aspect of design as co-evolving of problem-solution~\cite{dorst2001creativity}.} 
\rv{That is, what needs to be understood, such as what characteristics of model output are important, is emergent from the designs being explored, the depth of design details, and also the evolving understanding of the users' needs and characteristics.}
\rv{We suggest a few directions to support the contextualization and evolving information-seeking processes~\cite{kaur2022sensible,miller2019explanation,savolainen1993sense} of model understanding through \textit{model interrogation}. }
 
 %Participants were keen to look for examples that would be representative of articles shared on the Microblogging platform, hypothesized about impacting factors that would matter to the users, and translated model limitations to UX risks specific to their designs.
 
\rv{\textbf{\textit{Supporting example-based interrogation.}} Echoing prior research~\cite{subramonyam2021protoai,lee2022coauthor}, we found that examining input-output examples plays an important role in design decisions, as it allows designers to visually envision user reactions and design opportunities, as well as discover nuanced model behaviors that cannot be conveyed by high-level descriptions or metrics. However, an ad-hoc approach that relies on designers to choose the examples to examine does not guarantee an accurate and complete understanding, and can disadvantage inexperienced AI designers. Future work should explore helping designers create or customize example datasets that are representative of their use cases, and guiding them to explore the input and output spaces in a more systematic fashion, such as by suggesting examples from different categories. Designers should not only experience model affordances but also failures from examples, such as through observing edge cases. Lastly, it is necessary for designers to understand the generalizability of model behaviors they observe in examples, such as by having metrics quantifying their frequency and performing group-level output analyses that expand beyond basic disaggregated evaluations.}

\rv{ \textbf{\textit{Explainable AI for designerly understanding.}}} We note a potential role that explainable AI (XAI) tools can play in supporting designerly understanding, as ``how'' and ``why'' questions frequently emerged in participants' ideation processes. In human communication, people seek explanations about an event to be able to extrapolate to predictions about future, unseen events~\cite{lombrozo2006structure,lombrozo2012explanation}. This was often the goal and process that participants followed; they \rv{requested explanations of the output for} an example they observed, and \rv{then attempted to infer whether the model would behave similarly for other articles, and if so, what kinds. In some cases, participants also requested global explanations to infer general characteristics of the model's behavior and outputs.} 
\rv{\citet{subramonyam2022solving} found that when interacting with data scientists who had developed in-house models, designers often attempt to validate}
their hypotheses about model behaviors, errors, and impacting factors by seeking explanations about training data, features used, feature weights, rules, and underlying assumptions. Future work can explore how to provide similar \rv{information through} interactive \rv{explanations} for pre-trained models.

 \rv{\textbf{\textit{Supporting testing factors for UX impact.}}}  A recent development in making model reporting interrogatable is allowing users to provide or slice data to generate performance reports~\cite{goel2021robustness} \rv{for different groups}. Also available are a set of model behavioral analysis tools (e.g.,~\cite{wu2019errudite,ribeiro2020beyond,amershi2015modeltracker,chung2019slice}) to support ML engineers to understand impacting factors by performing disaggregated evaluation. However, their utility for UX designers is unclear. \rv{First, as discussed, UX failures may be concerned with a broader set of output characteristics than model errors. Second, designers face challenges translating between factors that impact model performance and the user scenarios they are designing for. For example, upon reading about the impact of an article being ``unstructured'' on performance, participants needed to translate that into ``what kind of users tend to share unstructured articles and what are the potential risks to them.'' When considering a common scenario of users sharing multiple articles in a row, P3 had to mentally simulate the output of chaining multiple articles. Future work should support designers to test factors of interest by allowing them to define metrics that matter to UX, and explore the impact by different user scenarios. For example, we may envision a tool that asks designers to define different user scenarios, helps them identify input examples that fit these scenarios, and allows visually examining their outputs and UX impact for each scenario. } 
 
\rv{\textbf{\textit{Integrating the exploration of design space and model understanding.}}  Recent HCI work begins to develop prototyping tools that integrate generating model outputs and creating interface designs in one place~\cite{subramonyam2021protoai}. Future work should explore AI prototyping tools that also incorporate model transparency information. Moreover, we observe that, to cope with the uncertainty and complexity of AI, there is a strong desire to create intermediate designs and explore how the model behaves for different design ideas, and assess the potential UX risks and benefits. Future tooling should support such processes and model understanding needs that emerge from exploring different designs. For example, P3's ideation in Figure~\ref{fig:designs} shows a natural inclination to create ``branching views'' to explore and manage different conditional designs, and their design decisions can benefit from a more concrete understanding of the model input and output characteristics for each branch of design.  }

\rv{Finally, we call out the immense need to support understanding large, general-purpose pre-trained models (e.g., GPT-3 and Dall-E) through interrogation to support designers or other individuals in making responsible decisions about their use. Given the extreme uncertainty about these models' capabilities and limitations, and the current uncertainty about appropriate application domains, any static documentation is unlikely to suffice. For example, the current documentation for GPT-3 provides only a high-level description of its capabilities, such as ``a set of models that can understand and generate natural language,'' and ``safety best practices guidance'' that includes examples of harms and mitigation strategies. We believe users of large pre-trained models can benefit from tools that support example-based interrogation, model behavior analysis on different input groups, risk-oriented explorations to discover context-specific failures and harms, capabilities to answer questions and test hypotheses, and tinkering with application ideas. }

\subsection{Implications for \rv{UX-Led Approaches for }RAI Practices }

\rv{Our study investigated designers' use of critical information about model limitations. Based on the results, we highlight a few reasons that UX practitioners can and should play a more central role in RAI practices to mitigate potential harms of AI technologies. First, UX design is fundamentally about bridging user needs and technical affordances. UX designers' training equips them with the skills to understand users through user research and prior experience, and extrapolate that understanding to anticipate user perceptions and behaviors interacting with a given technology. They can apply the same skills to anticipating potential harms of AI. For example, recently it has become more common to adopt ``red teaming'' into RAI development practices~\cite{ganguli2022red,fidel2012human}---coming up with adversarial inputs that produce harmful outputs and then updating the model to avoid them. We believe designers are well suited for performing such tasks.} Their bridging role also places a sense of responsibility on them to be ``user advocates,'' making them inclined to exhibit appropriate skepticism about a technology and actively seek to understand model limitations.

\rv{Second, as the four goals identified in Section~\ref{sec:goals} show, UX designers bring a unique set of tools to cope with limitations of AI and mitigate their potential harms. They are able to explore the design space, assess potential harms that are emergent from different design solutions, eliminate risky technology designs, and help identify harms-mitigation strategies that should drive technical development. They are able to create ``\textit{interaction-level interventions}'' (P9) to mitigate potential harms, by putting guardrails on model inputs and outputs, as well as creating transparency and user-control features to enable user agency dealing with model limitations.} While the AI ethics literature often asks the question of ``whether to build a technology''~\cite{barocas2020not}, UX practices and HCI literature have long contemplated with ``\textit{how} to build a technology'' responsibly when there are competing requirements and values (e.g., design for ``wicked problems''~\cite{buchanan1992wicked}, and applying it to RAI~\cite{park2022designing}). The core interest in supporting user agency is motivated by a critical position that even following the best practices, technology creators will always face uncertain downstream trajectories of use~\cite{inman2019beautiful,chalmers2004seamful,dix2007designing}. RAI practices can benefit from these user-centered and pragmatic design tools.

\rv{Lastly,} the increasing adoption of pre-trained models for product development \rv{calls for a UX-led approach as it necessitates design ideation to define new features that the model is suitable for. If well supported, pre-trained models also} present opportunities to empower UX practitioners to directly tinker with a wide range of AI design materials, prototype by choosing from and ``stitching together'' different design materials, and ultimately take a more proactive \rv{role in developing AI-powered products}.

\rv{With these arguments, we advocate for more and earlier inclusion of UX practitioners in the practice of RAI. This can be accomplished by providing designer-centered documentation and tooling to support their needs for understanding AI, as well as by lowering the organizational barriers for designers to take leading roles in product ideation, suitability assessment of models, and the definition of harms and mitigation strategies.   }

 \subsection{Limitations}
 As with any studies that utilize a \rv{design} probe, our results are contingent on the documentation we chose. However, our main focus on the categories of information \rv{required to meet designers' goals}, rather than specific content details, may mitigate this limitation to some degree. The extractive text summarization model and the design task also had their idiosyncrasies, and hence certain design processes may not generalize to other types of models and AI-powered features. While we aimed to introduce a hands-on task to observe participants' natural ideation processes, the study may still suffer from certain ecological validity issues. For example, participants were not given the opportunity to research the user problem, and the time and resources given were limited. Moreover, our sample was biased towards UX practitioners working in large technology companies and experienced with AI design. \rv{In fact, the majority of participants were recruited from a single large company. This sample may limit the generalizability of our observations about current practices reported in Section~\ref{sec:current}. For example, we mentioned that designers in smaller organizations appeared to take more initiative in the ideation stage. } Lastly, given that design ideation with pre-trained models is still an emerging task and not all our participants had engaged with such a task before, the design practices we observed may not cover all practices, and we do not claim that all observations should be taken as best practices. 
 
 \section{Conclusion}
 We conducted an interview study, including a hands-on design task, with 23 UX practitioners to investigate their needs and goals when performing design ideation on how to use a pre-trained ML model.  We took a primary interest in their information needs to develop a ``designerly understanding'' of the model, and explored whether and how information categories in transparent model reporting frameworks can support such an understanding. Our results inform future development of transparent model reporting practices, as well as other tools that aim to support design ideation working with pre-trained models.  Our study is motivated by two current trends in the broader context of AI product development: the availability of increasingly \rv{versatile} pre-trained models, including large, general-purpose pre-trained models, for product innovation; and the recognition of the importance and challenges of RAI practices that aim to proactively mitigate the potential harms, and safeguard the use of AI. We take a formative step towards exploring, and ultimately supporting, the opportunities and responsibility for UX practitioners under these trends.

\begin{acks}
We thank all our study participants and reviewers for their helpful feedback. We also thank Su Lin Blodgett, Alexandra Olteanu, Qiaosi Wang, and Ziang Xiao for their input and feedback on our study materials and procedure.  
\end{acks}

\bibliographystyle{ACM-Reference-Format}
\bibliography{refs}


\begin{thebibliography}{106}

%%% the Web address.

\ifx \showCODEN    \undefined \def \showCODEN     #1{\unskip}     \fi
\ifx \showDOI      \undefined \def \showDOI       #1{#1}\fi
\ifx \showISBNx    \undefined \def \showISBNx     #1{\unskip}     \fi
\ifx \showISBNxiii \undefined \def \showISBNxiii  #1{\unskip}     \fi
\ifx \showISSN     \undefined \def \showISSN      #1{\unskip}     \fi
\ifx \showLCCN     \undefined \def \showLCCN      #1{\unskip}     \fi
\ifx \shownote     \undefined \def \shownote      #1{#1}          \fi
\ifx \showarticletitle \undefined \def \showarticletitle #1{#1}   \fi
\ifx \showURL      \undefined \def \showURL       {\relax}        \fi
% invisible to TeX
\providecommand\bibfield[2]{#2}
\providecommand\bibinfo[2]{#2}
\providecommand\natexlab[1]{#1}
\providecommand\showeprint[2][]{arXiv:#2}

\bibitem[Ama(2019)]%
        {Amazon}
 \bibinfo{year}{2019}\natexlab{}.
\newblock \bibinfo{title}{Amazon AWS Machine Learning Services}.
\newblock
\newblock
\newblock
\shownote{https://aws.amazon.com/machine-learning/}.


\bibitem[goo(2019)]%
        {google-ai}
 \bibinfo{year}{2019}\natexlab{}.
\newblock \bibinfo{title}{Google AI for Developers}.
\newblock
\newblock
\newblock
\shownote{https://cloud.google.com/products/ai}.


\bibitem[pai(2019)]%
        {pair}
 \bibinfo{year}{2019}\natexlab{}.
\newblock \bibinfo{title}{Google People + AI Guidebook.}
\newblock
\newblock
\newblock
\shownote{pair.withgoogle.com/guidebook}.


\bibitem[hug(2019)]%
        {huggingface}
 \bibinfo{year}{2019}\natexlab{}.
\newblock \bibinfo{title}{Hugging Face Models}.
\newblock
\newblock
\newblock
\shownote{https://huggingface.co/models}.


\bibitem[IBM(2019)]%
        {IBM-AI}
 \bibinfo{year}{2019}\natexlab{}.
\newblock \bibinfo{title}{IBM Design for AI}.
\newblock
\newblock
\newblock
\shownote{https://www.ibm.com/design/ai/}.


\bibitem[wat(2019)]%
        {watson}
 \bibinfo{year}{2019}\natexlab{}.
\newblock \bibinfo{title}{IBM Watson AI Solutions}.
\newblock
\newblock
\newblock
\shownote{https://www.ibm.com/artificial-intelligence}.


\bibitem[Azu(2019)]%
        {Azure}
 \bibinfo{year}{2019}\natexlab{}.
\newblock \bibinfo{title}{Microsoft Azure Cognitive Service}.
\newblock
\newblock
\newblock
\shownote{https://azure.microsoft.com/en-us/services/cognitive-services/}.


\bibitem[tra(2022)]%
        {transparency-note}
 \bibinfo{year}{2022}\natexlab{}.
\newblock \bibinfo{title}{Transparency Note for Azure Cognitive Service for
  language}.
\newblock
\newblock
\newblock
\shownote{https://docs.microsoft.com/en-us/legal/cognitive-services/language-service/transparency-note}.


\bibitem[Allahyari et~al\mbox{.}(2017)]%
        {allahyari2017text}
\bibfield{author}{\bibinfo{person}{Mehdi Allahyari}, \bibinfo{person}{Seyedamin
  Pouriyeh}, \bibinfo{person}{Mehdi Assefi}, \bibinfo{person}{Saeid Safaei},
  \bibinfo{person}{Elizabeth~D Trippe}, \bibinfo{person}{Juan~B Gutierrez},
  {and} \bibinfo{person}{Krys Kochut}.} \bibinfo{year}{2017}\natexlab{}.
\newblock \showarticletitle{Text summarization techniques: a brief survey}.
\newblock \bibinfo{journal}{\emph{arXiv preprint arXiv:1707.02268}}
  (\bibinfo{year}{2017}).
\newblock


\bibitem[Amershi et~al\mbox{.}(2015)]%
        {amershi2015modeltracker}
\bibfield{author}{\bibinfo{person}{Saleema Amershi}, \bibinfo{person}{Max
  Chickering}, \bibinfo{person}{Steven~M Drucker}, \bibinfo{person}{Bongshin
  Lee}, \bibinfo{person}{Patrice Simard}, {and} \bibinfo{person}{Jina Suh}.}
  \bibinfo{year}{2015}\natexlab{}.
\newblock \showarticletitle{Modeltracker: Redesigning performance analysis
  tools for machine learning}. In \bibinfo{booktitle}{\emph{Proceedings of the
  33rd Annual ACM Conference on Human Factors in Computing Systems}}.
  \bibinfo{pages}{337--346}.
\newblock


\bibitem[Amershi et~al\mbox{.}(2019)]%
        {amershi2019guidelines}
\bibfield{author}{\bibinfo{person}{Saleema Amershi}, \bibinfo{person}{Dan
  Weld}, \bibinfo{person}{Mihaela Vorvoreanu}, \bibinfo{person}{Adam Fourney},
  \bibinfo{person}{Besmira Nushi}, \bibinfo{person}{Penny Collisson},
  \bibinfo{person}{Jina Suh}, \bibinfo{person}{Shamsi Iqbal},
  \bibinfo{person}{Paul~N Bennett}, \bibinfo{person}{Kori Inkpen},
  {et~al\mbox{.}}} \bibinfo{year}{2019}\natexlab{}.
\newblock \showarticletitle{Guidelines for human-AI interaction}. In
  \bibinfo{booktitle}{\emph{Proceedings of the 2019 chi conference on human
  factors in computing systems}}. \bibinfo{pages}{1--13}.
\newblock


\bibitem[Arnold et~al\mbox{.}(2019)]%
        {arnold2019factsheets}
\bibfield{author}{\bibinfo{person}{Matthew Arnold}, \bibinfo{person}{Rachel~KE
  Bellamy}, \bibinfo{person}{Michael Hind}, \bibinfo{person}{Stephanie Houde},
  \bibinfo{person}{Sameep Mehta}, \bibinfo{person}{Aleksandra Mojsilovi{\'c}},
  \bibinfo{person}{Ravi Nair}, \bibinfo{person}{K~Natesan Ramamurthy},
  \bibinfo{person}{Alexandra Olteanu}, \bibinfo{person}{David Piorkowski},
  {et~al\mbox{.}}} \bibinfo{year}{2019}\natexlab{}.
\newblock \showarticletitle{FactSheets: Increasing trust in AI services through
  supplier's declarations of conformity}.
\newblock \bibinfo{journal}{\emph{IBM Journal of Research and Development}}
  \bibinfo{volume}{63}, \bibinfo{number}{4/5} (\bibinfo{year}{2019}),
  \bibinfo{pages}{6--1}.
\newblock


\bibitem[Barocas et~al\mbox{.}(2020)]%
        {barocas2020not}
\bibfield{author}{\bibinfo{person}{Solon Barocas}, \bibinfo{person}{Asia~J
  Biega}, \bibinfo{person}{Benjamin Fish}, \bibinfo{person}{J{\k{e}}drzej
  Niklas}, {and} \bibinfo{person}{Luke Stark}.}
  \bibinfo{year}{2020}\natexlab{}.
\newblock \showarticletitle{When not to design, build, or deploy}. In
  \bibinfo{booktitle}{\emph{Proceedings of the 2020 Conference on Fairness,
  Accountability, and Transparency}}. \bibinfo{pages}{695--695}.
\newblock


\bibitem[Barocas et~al\mbox{.}(2021)]%
        {barocas2021designing}
\bibfield{author}{\bibinfo{person}{Solon Barocas}, \bibinfo{person}{Anhong
  Guo}, \bibinfo{person}{Ece Kamar}, \bibinfo{person}{Jacquelyn Krones},
  \bibinfo{person}{Meredith~Ringel Morris}, \bibinfo{person}{Jennifer~Wortman
  Vaughan}, \bibinfo{person}{W~Duncan Wadsworth}, {and} \bibinfo{person}{Hanna
  Wallach}.} \bibinfo{year}{2021}\natexlab{}.
\newblock \showarticletitle{Designing disaggregated evaluations of ai systems:
  Choices, considerations, and tradeoffs}. In
  \bibinfo{booktitle}{\emph{Proceedings of the 2021 AAAI/ACM Conference on AI,
  Ethics, and Society}}. \bibinfo{pages}{368--378}.
\newblock


\bibitem[Bender and Friedman(2018)]%
        {bender2018data}
\bibfield{author}{\bibinfo{person}{Emily~M. Bender} {and}
  \bibinfo{person}{Batya Friedman}.} \bibinfo{year}{2018}\natexlab{}.
\newblock \showarticletitle{Data Statements for Natural Language Processing:
  Toward Mitigating System Bias and Enabling Better Science}.
\newblock \bibinfo{journal}{\emph{Transactions of the Association for
  Computational Linguistics}}  \bibinfo{volume}{6} (\bibinfo{year}{2018}),
  \bibinfo{pages}{587--604}.
\newblock


\bibitem[Benjamin et~al\mbox{.}(2021)]%
        {benjamin2021machine}
\bibfield{author}{\bibinfo{person}{Jesse~Josua Benjamin}, \bibinfo{person}{Arne
  Berger}, \bibinfo{person}{Nick Merrill}, {and} \bibinfo{person}{James
  Pierce}.} \bibinfo{year}{2021}\natexlab{}.
\newblock \showarticletitle{Machine Learning Uncertainty as a Design Material:
  A Post-Phenomenological Inquiry}. In \bibinfo{booktitle}{\emph{Proceedings of
  the 2021 CHI Conference on Human Factors in Computing Systems}}.
  \bibinfo{pages}{1--14}.
\newblock


\bibitem[Benjamins et~al\mbox{.}(2019)]%
        {benjamins2019responsible}
\bibfield{author}{\bibinfo{person}{Richard Benjamins}, \bibinfo{person}{Alberto
  Barbado}, {and} \bibinfo{person}{Daniel Sierra}.}
  \bibinfo{year}{2019}\natexlab{}.
\newblock \showarticletitle{Responsible AI by design in practice}.
\newblock \bibinfo{journal}{\emph{arXiv preprint arXiv:1909.12838}}
  (\bibinfo{year}{2019}).
\newblock


\bibitem[Bhatt et~al\mbox{.}(2020)]%
        {bhatt2020explainable}
\bibfield{author}{\bibinfo{person}{Umang Bhatt}, \bibinfo{person}{Alice Xiang},
  \bibinfo{person}{Shubham Sharma}, \bibinfo{person}{Adrian Weller},
  \bibinfo{person}{Ankur Taly}, \bibinfo{person}{Yunhan Jia},
  \bibinfo{person}{Joydeep Ghosh}, \bibinfo{person}{Ruchir Puri},
  \bibinfo{person}{Jos{\'e}~MF Moura}, {and} \bibinfo{person}{Peter
  Eckersley}.} \bibinfo{year}{2020}\natexlab{}.
\newblock \showarticletitle{Explainable machine learning in deployment}. In
  \bibinfo{booktitle}{\emph{Proceedings of the 2020 conference on fairness,
  accountability, and transparency}}. \bibinfo{pages}{648--657}.
\newblock


\bibitem[Blodgett et~al\mbox{.}(2020)]%
        {blodgett2020language}
\bibfield{author}{\bibinfo{person}{Su~Lin Blodgett}, \bibinfo{person}{Solon
  Barocas}, \bibinfo{person}{Hal Daum{\'e}~III}, {and} \bibinfo{person}{Hanna
  Wallach}.} \bibinfo{year}{2020}\natexlab{}.
\newblock \showarticletitle{Language (Technology) is Power: A Critical Survey
  of “Bias” in NLP}. In \bibinfo{booktitle}{\emph{Proceedings of the 58th
  Annual Meeting of the Association for Computational Linguistics}}.
  \bibinfo{pages}{5454--5476}.
\newblock


\bibitem[Boyd(2021)]%
        {boyd2021datasheets}
\bibfield{author}{\bibinfo{person}{Karen~L Boyd}.}
  \bibinfo{year}{2021}\natexlab{}.
\newblock \showarticletitle{Datasheets for Datasets help ML Engineers Notice
  and Understand Ethical Issues in Training Data}.
\newblock \bibinfo{journal}{\emph{Proceedings of the ACM on Human-Computer
  Interaction}} \bibinfo{volume}{5}, \bibinfo{number}{CSCW2}
  (\bibinfo{year}{2021}), \bibinfo{pages}{1--27}.
\newblock


\bibitem[Brown et~al\mbox{.}(2019)]%
        {brown2019toward}
\bibfield{author}{\bibinfo{person}{Anna Brown}, \bibinfo{person}{Alexandra
  Chouldechova}, \bibinfo{person}{Emily Putnam-Hornstein},
  \bibinfo{person}{Andrew Tobin}, {and} \bibinfo{person}{Rhema Vaithianathan}.}
  \bibinfo{year}{2019}\natexlab{}.
\newblock \showarticletitle{Toward algorithmic accountability in public
  services: A qualitative study of affected community perspectives on
  algorithmic decision-making in child welfare services}. In
  \bibinfo{booktitle}{\emph{Proceedings of the 2019 CHI Conference on Human
  Factors in Computing Systems}}. \bibinfo{pages}{1--12}.
\newblock


\bibitem[Brown et~al\mbox{.}(2020)]%
        {brown2020language}
\bibfield{author}{\bibinfo{person}{Tom Brown}, \bibinfo{person}{Benjamin Mann},
  \bibinfo{person}{Nick Ryder}, \bibinfo{person}{Melanie Subbiah},
  \bibinfo{person}{Jared~D Kaplan}, \bibinfo{person}{Prafulla Dhariwal},
  \bibinfo{person}{Arvind Neelakantan}, \bibinfo{person}{Pranav Shyam},
  \bibinfo{person}{Girish Sastry}, \bibinfo{person}{Amanda Askell},
  {et~al\mbox{.}}} \bibinfo{year}{2020}\natexlab{}.
\newblock \showarticletitle{Language models are few-shot learners}.
\newblock \bibinfo{journal}{\emph{Advances in neural information processing
  systems}}  \bibinfo{volume}{33} (\bibinfo{year}{2020}),
  \bibinfo{pages}{1877--1901}.
\newblock


\bibitem[Buchanan(1992)]%
        {buchanan1992wicked}
\bibfield{author}{\bibinfo{person}{Richard Buchanan}.}
  \bibinfo{year}{1992}\natexlab{}.
\newblock \showarticletitle{Wicked problems in design thinking}.
\newblock \bibinfo{journal}{\emph{Design issues}} \bibinfo{volume}{8},
  \bibinfo{number}{2} (\bibinfo{year}{1992}), \bibinfo{pages}{5--21}.
\newblock


\bibitem[Buxton(2010)]%
        {buxton2010sketching}
\bibfield{author}{\bibinfo{person}{Bill Buxton}.}
  \bibinfo{year}{2010}\natexlab{}.
\newblock \bibinfo{booktitle}{\emph{Sketching user experiences: getting the
  design right and the right design}}.
\newblock \bibinfo{publisher}{Morgan kaufmann}.
\newblock


\bibitem[Cai et~al\mbox{.}(2021)]%
        {cai2021onboarding}
\bibfield{author}{\bibinfo{person}{Carrie~J Cai}, \bibinfo{person}{Samantha
  Winter}, \bibinfo{person}{David Steiner}, \bibinfo{person}{Lauren Wilcox},
  {and} \bibinfo{person}{Michael Terry}.} \bibinfo{year}{2021}\natexlab{}.
\newblock \showarticletitle{Onboarding Materials as Cross-functional Boundary
  Objects for Developing AI Assistants}. In \bibinfo{booktitle}{\emph{Extended
  Abstracts of the 2021 CHI Conference on Human Factors in Computing Systems}}.
  \bibinfo{pages}{1--7}.
\newblock


\bibitem[Chalmers and Galani(2004)]%
        {chalmers2004seamful}
\bibfield{author}{\bibinfo{person}{Matthew Chalmers} {and}
  \bibinfo{person}{Areti Galani}.} \bibinfo{year}{2004}\natexlab{}.
\newblock \showarticletitle{Seamful interweaving: heterogeneity in the theory
  and design of interactive systems}. In \bibinfo{booktitle}{\emph{Proceedings
  of the 5th conference on Designing interactive systems: processes, practices,
  methods, and techniques}}. \bibinfo{pages}{243--252}.
\newblock


\bibitem[Chen et~al\mbox{.}(2022)]%
        {chen2022interpretable}
\bibfield{author}{\bibinfo{person}{Valerie Chen}, \bibinfo{person}{Jeffrey Li},
  \bibinfo{person}{Joon~Sik Kim}, \bibinfo{person}{Gregory Plumb}, {and}
  \bibinfo{person}{Ameet Talwalkar}.} \bibinfo{year}{2022}\natexlab{}.
\newblock \showarticletitle{Interpretable machine learning: Moving from mythos
  to diagnostics}.
\newblock \bibinfo{journal}{\emph{Queue}} \bibinfo{volume}{19},
  \bibinfo{number}{6} (\bibinfo{year}{2022}), \bibinfo{pages}{28--56}.
\newblock


\bibitem[Chung et~al\mbox{.}(2019)]%
        {chung2019slice}
\bibfield{author}{\bibinfo{person}{Yeounoh Chung}, \bibinfo{person}{Tim
  Kraska}, \bibinfo{person}{Neoklis Polyzotis}, \bibinfo{person}{Ki~Hyun Tae},
  {and} \bibinfo{person}{Steven~Euijong Whang}.}
  \bibinfo{year}{2019}\natexlab{}.
\newblock \showarticletitle{Slice finder: Automated data slicing for model
  validation}. In \bibinfo{booktitle}{\emph{2019 IEEE 35th International
  Conference on Data Engineering (ICDE)}}. IEEE, \bibinfo{pages}{1550--1553}.
\newblock


\bibitem[Corbin et~al\mbox{.}(2015)]%
        {corbin2015basics}
\bibfield{author}{\bibinfo{person}{Juliet Corbin}, \bibinfo{person}{Anselm~L
  Strauss}, {and} \bibinfo{person}{Anselm Strauss}.}
  \bibinfo{year}{2015}\natexlab{}.
\newblock \bibinfo{booktitle}{\emph{Basics of qualitative research}}.
\newblock \bibinfo{publisher}{sage}.
\newblock


\bibitem[Council(2005)]%
        {diamond}
\bibfield{author}{\bibinfo{person}{Design Council}.}
  \bibinfo{year}{2005}\natexlab{}.
\newblock \showarticletitle{The ‘double diamond’ design process model.
  Design Counci}.
\newblock  (\bibinfo{year}{2005}).
\newblock


\bibitem[Cramer and Kim(2019)]%
        {cramer2019confronting}
\bibfield{author}{\bibinfo{person}{Henriette Cramer} {and}
  \bibinfo{person}{Juho Kim}.} \bibinfo{year}{2019}\natexlab{}.
\newblock \showarticletitle{Confronting the tensions where UX meets AI}.
\newblock \bibinfo{journal}{\emph{Interactions}} \bibinfo{volume}{26},
  \bibinfo{number}{6} (\bibinfo{year}{2019}), \bibinfo{pages}{69--71}.
\newblock


\bibitem[Crisan et~al\mbox{.}(2022)]%
        {crisan2022interactive}
\bibfield{author}{\bibinfo{person}{Anamaria Crisan}, \bibinfo{person}{Margaret
  Drouhard}, \bibinfo{person}{Jesse Vig}, {and} \bibinfo{person}{Nazneen
  Rajani}.} \bibinfo{year}{2022}\natexlab{}.
\newblock \showarticletitle{Interactive Model Cards: A Human-Centered Approach
  to Model Documentation}.
\newblock \bibinfo{journal}{\emph{arXiv preprint arXiv:2205.02894}}
  (\bibinfo{year}{2022}).
\newblock


\bibitem[Delgado et~al\mbox{.}(2021)]%
        {delgado2021stakeholder}
\bibfield{author}{\bibinfo{person}{Fernando Delgado}, \bibinfo{person}{Stephen
  Yang}, \bibinfo{person}{Michael Madaio}, {and} \bibinfo{person}{Qian Yang}.}
  \bibinfo{year}{2021}\natexlab{}.
\newblock \showarticletitle{Stakeholder Participation in AI: Beyond" Add
  Diverse Stakeholders and Stir"}.
\newblock \bibinfo{journal}{\emph{arXiv preprint arXiv:2111.01122}}
  (\bibinfo{year}{2021}).
\newblock


\bibitem[Dellermann et~al\mbox{.}(2021)]%
        {dellermann2021future}
\bibfield{author}{\bibinfo{person}{Dominik Dellermann}, \bibinfo{person}{Adrian
  Calma}, \bibinfo{person}{Nikolaus Lipusch}, \bibinfo{person}{Thorsten Weber},
  \bibinfo{person}{Sascha Weigel}, {and} \bibinfo{person}{Philipp Ebel}.}
  \bibinfo{year}{2021}\natexlab{}.
\newblock \showarticletitle{The future of human-AI collaboration: a taxonomy of
  design knowledge for hybrid intelligence systems}.
\newblock \bibinfo{journal}{\emph{arXiv preprint arXiv:2105.03354}}
  (\bibinfo{year}{2021}).
\newblock


\bibitem[Deng et~al\mbox{.}(2022)]%
        {deng2022exploring}
\bibfield{author}{\bibinfo{person}{Wesley~Hanwen Deng}, \bibinfo{person}{Manish
  Nagireddy}, \bibinfo{person}{Michelle Seng~Ah Lee}, \bibinfo{person}{Jatinder
  Singh}, \bibinfo{person}{Zhiwei~Steven Wu}, \bibinfo{person}{Kenneth
  Holstein}, {and} \bibinfo{person}{Haiyi Zhu}.}
  \bibinfo{year}{2022}\natexlab{}.
\newblock \showarticletitle{Exploring How Machine Learning Practitioners (Try
  To) Use Fairness Toolkits}.
\newblock \bibinfo{journal}{\emph{arXiv preprint arXiv:2205.06922}}
  (\bibinfo{year}{2022}).
\newblock


\bibitem[Dix(2007)]%
        {dix2007designing}
\bibfield{author}{\bibinfo{person}{Alan Dix}.} \bibinfo{year}{2007}\natexlab{}.
\newblock \showarticletitle{Designing for appropriation}. In
  \bibinfo{booktitle}{\emph{Proceedings of HCI 2007 The 21st British HCI Group
  Annual Conference University of Lancaster, UK 21}}. \bibinfo{pages}{1--4}.
\newblock


\bibitem[Doordan(2003)]%
        {doordan2003materials}
\bibfield{author}{\bibinfo{person}{Dennis~P Doordan}.}
  \bibinfo{year}{2003}\natexlab{}.
\newblock \showarticletitle{On materials}.
\newblock \bibinfo{journal}{\emph{Design Issues}} \bibinfo{volume}{19},
  \bibinfo{number}{4} (\bibinfo{year}{2003}), \bibinfo{pages}{3--8}.
\newblock


\bibitem[Dorst and Cross(2001)]%
        {dorst2001creativity}
\bibfield{author}{\bibinfo{person}{Kees Dorst} {and} \bibinfo{person}{Nigel
  Cross}.} \bibinfo{year}{2001}\natexlab{}.
\newblock \showarticletitle{Creativity in the design process: co-evolution of
  problem--solution}.
\newblock \bibinfo{journal}{\emph{Design studies}} \bibinfo{volume}{22},
  \bibinfo{number}{5} (\bibinfo{year}{2001}), \bibinfo{pages}{425--437}.
\newblock


\bibitem[Dove et~al\mbox{.}(2017)]%
        {dove2017ux}
\bibfield{author}{\bibinfo{person}{Graham Dove}, \bibinfo{person}{Kim Halskov},
  \bibinfo{person}{Jodi Forlizzi}, {and} \bibinfo{person}{John Zimmerman}.}
  \bibinfo{year}{2017}\natexlab{}.
\newblock \showarticletitle{UX design innovation: Challenges for working with
  machine learning as a design material}. In
  \bibinfo{booktitle}{\emph{Proceedings of the 2017 chi conference on human
  factors in computing systems}}. \bibinfo{pages}{278--288}.
\newblock


\bibitem[Eiband et~al\mbox{.}(2018)]%
        {eiband2018bringing}
\bibfield{author}{\bibinfo{person}{Malin Eiband}, \bibinfo{person}{Hanna
  Schneider}, \bibinfo{person}{Mark Bilandzic}, \bibinfo{person}{Julian
  Fazekas-Con}, \bibinfo{person}{Mareike Haug}, {and} \bibinfo{person}{Heinrich
  Hussmann}.} \bibinfo{year}{2018}\natexlab{}.
\newblock \showarticletitle{Bringing transparency design into practice}. In
  \bibinfo{booktitle}{\emph{23rd international conference on intelligent user
  interfaces}}. \bibinfo{pages}{211--223}.
\newblock


\bibitem[Feng et~al\mbox{.}(2021)]%
        {feng2021survey}
\bibfield{author}{\bibinfo{person}{Xiachong Feng}, \bibinfo{person}{Xiaocheng
  Feng}, {and} \bibinfo{person}{Bing Qin}.} \bibinfo{year}{2021}\natexlab{}.
\newblock \showarticletitle{A survey on dialogue summarization: Recent advances
  and new frontiers}.
\newblock \bibinfo{journal}{\emph{arXiv preprint arXiv:2107.03175}}
  (\bibinfo{year}{2021}).
\newblock


\bibitem[Fernaeus and Sundstr{\"o}m(2012)]%
        {fernaeus2012material}
\bibfield{author}{\bibinfo{person}{Ylva Fernaeus} {and} \bibinfo{person}{Petra
  Sundstr{\"o}m}.} \bibinfo{year}{2012}\natexlab{}.
\newblock \showarticletitle{The material move how materials matter in
  interaction design research}. In \bibinfo{booktitle}{\emph{proceedings of the
  designing interactive systems conference}}. \bibinfo{pages}{486--495}.
\newblock


\bibitem[Fidel(2012)]%
        {fidel2012human}
\bibfield{author}{\bibinfo{person}{Raya Fidel}.}
  \bibinfo{year}{2012}\natexlab{}.
\newblock \bibinfo{booktitle}{\emph{Human information interaction: An
  ecological approach to information behavior}}.
\newblock \bibinfo{publisher}{Mit Press}.
\newblock


\bibitem[Ganguli et~al\mbox{.}(2022)]%
        {ganguli2022red}
\bibfield{author}{\bibinfo{person}{Deep Ganguli}, \bibinfo{person}{Liane
  Lovitt}, \bibinfo{person}{Jackson Kernion}, \bibinfo{person}{Amanda Askell},
  \bibinfo{person}{Yuntao Bai}, \bibinfo{person}{Saurav Kadavath},
  \bibinfo{person}{Ben Mann}, \bibinfo{person}{Ethan Perez},
  \bibinfo{person}{Nicholas Schiefer}, \bibinfo{person}{Kamal Ndousse},
  {et~al\mbox{.}}} \bibinfo{year}{2022}\natexlab{}.
\newblock \showarticletitle{Red teaming language models to reduce harms:
  Methods, scaling behaviors, and lessons learned}.
\newblock \bibinfo{journal}{\emph{arXiv preprint arXiv:2209.07858}}
  (\bibinfo{year}{2022}).
\newblock


\bibitem[Gebru et~al\mbox{.}(2021)]%
        {gebru2021datasheets}
\bibfield{author}{\bibinfo{person}{Timnit Gebru}, \bibinfo{person}{Jamie
  Morgenstern}, \bibinfo{person}{Briana Vecchione},
  \bibinfo{person}{Jennifer~Wortman Vaughan}, \bibinfo{person}{Hanna Wallach},
  \bibinfo{person}{Hal~Daum{\'e} Iii}, {and} \bibinfo{person}{Kate Crawford}.}
  \bibinfo{year}{2021}\natexlab{}.
\newblock \showarticletitle{Datasheets for datasets}.
\newblock \bibinfo{journal}{\emph{Commun. ACM}} \bibinfo{volume}{64},
  \bibinfo{number}{12} (\bibinfo{year}{2021}), \bibinfo{pages}{86--92}.
\newblock


\bibitem[Giaccardi and Karana(2015)]%
        {giaccardi2015foundations}
\bibfield{author}{\bibinfo{person}{Elisa Giaccardi} {and}
  \bibinfo{person}{Elvin Karana}.} \bibinfo{year}{2015}\natexlab{}.
\newblock \showarticletitle{Foundations of materials experience: An approach
  for HCI}. In \bibinfo{booktitle}{\emph{Proceedings of the 33rd Annual ACM
  Conference on Human Factors in Computing Systems}}.
  \bibinfo{pages}{2447--2456}.
\newblock


\bibitem[Goel et~al\mbox{.}(2021)]%
        {goel2021robustness}
\bibfield{author}{\bibinfo{person}{Karan Goel}, \bibinfo{person}{Nazneen
  Rajani}, \bibinfo{person}{Jesse Vig}, \bibinfo{person}{Samson Tan},
  \bibinfo{person}{Jason Wu}, \bibinfo{person}{Stephan Zheng},
  \bibinfo{person}{Caiming Xiong}, \bibinfo{person}{Mohit Bansal}, {and}
  \bibinfo{person}{Christopher R{\'e}}.} \bibinfo{year}{2021}\natexlab{}.
\newblock \showarticletitle{Robustness gym: Unifying the nlp evaluation
  landscape}.
\newblock \bibinfo{journal}{\emph{arXiv preprint arXiv:2101.04840}}
  (\bibinfo{year}{2021}).
\newblock


\bibitem[Guidotti et~al\mbox{.}(2018)]%
        {guidotti2018survey}
\bibfield{author}{\bibinfo{person}{Riccardo Guidotti}, \bibinfo{person}{Anna
  Monreale}, \bibinfo{person}{Salvatore Ruggieri}, \bibinfo{person}{Franco
  Turini}, \bibinfo{person}{Fosca Giannotti}, {and} \bibinfo{person}{Dino
  Pedreschi}.} \bibinfo{year}{2018}\natexlab{}.
\newblock \showarticletitle{A survey of methods for explaining black box
  models}.
\newblock \bibinfo{journal}{\emph{ACM computing surveys (CSUR)}}
  \bibinfo{volume}{51}, \bibinfo{number}{5} (\bibinfo{year}{2018}),
  \bibinfo{pages}{1--42}.
\newblock


\bibitem[Gunning et~al\mbox{.}(2019)]%
        {gunning2019xai}
\bibfield{author}{\bibinfo{person}{David Gunning}, \bibinfo{person}{Mark
  Stefik}, \bibinfo{person}{Jaesik Choi}, \bibinfo{person}{Timothy Miller},
  \bibinfo{person}{Simone Stumpf}, {and} \bibinfo{person}{Guang-Zhong Yang}.}
  \bibinfo{year}{2019}\natexlab{}.
\newblock \showarticletitle{XAI—Explainable artificial intelligence}.
\newblock \bibinfo{journal}{\emph{Science Robotics}} \bibinfo{volume}{4},
  \bibinfo{number}{37} (\bibinfo{year}{2019}).
\newblock


\bibitem[Heger et~al\mbox{.}(2022)]%
        {heger2022understanding}
\bibfield{author}{\bibinfo{person}{Amy~K. Heger}, \bibinfo{person}{Liz~B.
  Marquis}, \bibinfo{person}{Mihaela Vorvoreanu}, \bibinfo{person}{Hanna
  Wallach}, {and} \bibinfo{person}{Jennifer~Wortman Vaughan}.}
  \bibinfo{year}{2022}\natexlab{}.
\newblock \showarticletitle{Understanding Machine Learning Practitioners' Data
  Documentation Perceptions, Needs, Challenges, and Desiderata}.
\newblock \bibinfo{journal}{\emph{Proceedings of the ACM on Human-Computer
  Interaction}} \bibinfo{volume}{6}, \bibinfo{number}{CSCW2}
  (\bibinfo{year}{2022}).
\newblock


\bibitem[Hind et~al\mbox{.}(2020)]%
        {hind2020experiences}
\bibfield{author}{\bibinfo{person}{Michael Hind}, \bibinfo{person}{Stephanie
  Houde}, \bibinfo{person}{Jacquelyn Martino}, \bibinfo{person}{Aleksandra
  Mojsilovic}, \bibinfo{person}{David Piorkowski}, \bibinfo{person}{John
  Richards}, {and} \bibinfo{person}{Kush~R Varshney}.}
  \bibinfo{year}{2020}\natexlab{}.
\newblock \showarticletitle{Experiences with improving the transparency of AI
  models and services}. In \bibinfo{booktitle}{\emph{Extended Abstracts of the
  2020 CHI Conference on Human Factors in Computing Systems}}.
  \bibinfo{pages}{1--8}.
\newblock


\bibitem[Hohman et~al\mbox{.}(2019)]%
        {hohman2019gamut}
\bibfield{author}{\bibinfo{person}{Fred Hohman}, \bibinfo{person}{Andrew Head},
  \bibinfo{person}{Rich Caruana}, \bibinfo{person}{Robert DeLine}, {and}
  \bibinfo{person}{Steven~M Drucker}.} \bibinfo{year}{2019}\natexlab{}.
\newblock \showarticletitle{Gamut: A design probe to understand how data
  scientists understand machine learning models}. In
  \bibinfo{booktitle}{\emph{Proceedings of the 2019 CHI conference on human
  factors in computing systems}}. \bibinfo{pages}{1--13}.
\newblock


\bibitem[Holland et~al\mbox{.}(2020)]%
        {holland2020dataset}
\bibfield{author}{\bibinfo{person}{Sarah Holland}, \bibinfo{person}{Ahmed
  Hosny}, \bibinfo{person}{Sarah Newman}, \bibinfo{person}{Joshua Joseph},
  {and} \bibinfo{person}{Kasia Chmielinski}.} \bibinfo{year}{2020}\natexlab{}.
\newblock \showarticletitle{The dataset nutrition label}.
\newblock \bibinfo{journal}{\emph{Data Protection and Privacy, Volume 12: Data
  Protection and Democracy}}  \bibinfo{volume}{12} (\bibinfo{year}{2020}),
  \bibinfo{pages}{1}.
\newblock


\bibitem[Holmquist(2017)]%
        {holmquist2017intelligence}
\bibfield{author}{\bibinfo{person}{Lars~Erik Holmquist}.}
  \bibinfo{year}{2017}\natexlab{}.
\newblock \showarticletitle{Intelligence on tap: artificial intelligence as a
  new design material}.
\newblock \bibinfo{journal}{\emph{interactions}} \bibinfo{volume}{24},
  \bibinfo{number}{4} (\bibinfo{year}{2017}), \bibinfo{pages}{28--33}.
\newblock


\bibitem[Holstein et~al\mbox{.}(2019)]%
        {holstein2019improving}
\bibfield{author}{\bibinfo{person}{Kenneth Holstein}, \bibinfo{person}{Jennifer
  Wortman~Vaughan}, \bibinfo{person}{Hal Daum{\'e}~III}, \bibinfo{person}{Miro
  Dudik}, {and} \bibinfo{person}{Hanna Wallach}.}
  \bibinfo{year}{2019}\natexlab{}.
\newblock \showarticletitle{Improving fairness in machine learning systems:
  What do industry practitioners need?}. In
  \bibinfo{booktitle}{\emph{Proceedings of the 2019 CHI conference on human
  factors in computing systems}}. \bibinfo{pages}{1--16}.
\newblock


\bibitem[Hong et~al\mbox{.}(2021)]%
        {hong2021planning}
\bibfield{author}{\bibinfo{person}{Matthew~K Hong}, \bibinfo{person}{Adam
  Fourney}, \bibinfo{person}{Derek DeBellis}, {and} \bibinfo{person}{Saleema
  Amershi}.} \bibinfo{year}{2021}\natexlab{}.
\newblock \showarticletitle{Planning for natural language failures with the ai
  playbook}. In \bibinfo{booktitle}{\emph{Proceedings of the 2021 CHI
  Conference on Human Factors in Computing Systems}}. \bibinfo{pages}{1--11}.
\newblock


\bibitem[Hong et~al\mbox{.}(2020)]%
        {hong2020human}
\bibfield{author}{\bibinfo{person}{Sungsoo~Ray Hong}, \bibinfo{person}{Jessica
  Hullman}, {and} \bibinfo{person}{Enrico Bertini}.}
  \bibinfo{year}{2020}\natexlab{}.
\newblock \showarticletitle{Human factors in model interpretability: Industry
  practices, challenges, and needs}.
\newblock \bibinfo{journal}{\emph{Proceedings of the ACM on Human-Computer
  Interaction}} \bibinfo{volume}{4}, \bibinfo{number}{CSCW1}
  (\bibinfo{year}{2020}), \bibinfo{pages}{1--26}.
\newblock


\bibitem[Hutchinson et~al\mbox{.}(2003)]%
        {hutchinson2003technology}
\bibfield{author}{\bibinfo{person}{Hilary Hutchinson}, \bibinfo{person}{Wendy
  Mackay}, \bibinfo{person}{Bo Westerlund}, \bibinfo{person}{Benjamin~B
  Bederson}, \bibinfo{person}{Allison Druin}, \bibinfo{person}{Catherine
  Plaisant}, \bibinfo{person}{Michel Beaudouin-Lafon},
  \bibinfo{person}{St{\'e}phane Conversy}, \bibinfo{person}{Helen Evans},
  \bibinfo{person}{Heiko Hansen}, {et~al\mbox{.}}}
  \bibinfo{year}{2003}\natexlab{}.
\newblock \showarticletitle{Technology probes: inspiring design for and with
  families}. In \bibinfo{booktitle}{\emph{Proceedings of the SIGCHI conference
  on Human factors in computing systems}}. \bibinfo{pages}{17--24}.
\newblock


\bibitem[Inman and Ribes(2019)]%
        {inman2019beautiful}
\bibfield{author}{\bibinfo{person}{Sarah Inman} {and} \bibinfo{person}{David
  Ribes}.} \bibinfo{year}{2019}\natexlab{}.
\newblock \showarticletitle{" Beautiful Seams" Strategic Revelations and
  Concealments}. In \bibinfo{booktitle}{\emph{Proceedings of the 2019 CHI
  Conference on Human Factors in Computing Systems}}. \bibinfo{pages}{1--14}.
\newblock


\bibitem[Kaur et~al\mbox{.}(2022)]%
        {kaur2022sensible}
\bibfield{author}{\bibinfo{person}{Harmanpreet Kaur}, \bibinfo{person}{Eytan
  Adar}, \bibinfo{person}{Eric Gilbert}, {and} \bibinfo{person}{Cliff Lampe}.}
  \bibinfo{year}{2022}\natexlab{}.
\newblock \showarticletitle{Sensible AI: Re-imagining Interpretability and
  Explainability using Sensemaking Theory}.
\newblock \bibinfo{journal}{\emph{arXiv preprint arXiv:2205.05057}}
  (\bibinfo{year}{2022}).
\newblock


\bibitem[Kaur et~al\mbox{.}(2020)]%
        {kaur2020interpreting}
\bibfield{author}{\bibinfo{person}{Harmanpreet Kaur}, \bibinfo{person}{Harsha
  Nori}, \bibinfo{person}{Samuel Jenkins}, \bibinfo{person}{Rich Caruana},
  \bibinfo{person}{Hanna Wallach}, {and} \bibinfo{person}{Jennifer
  Wortman~Vaughan}.} \bibinfo{year}{2020}\natexlab{}.
\newblock \showarticletitle{Interpreting Interpretability: Understanding Data
  Scientists' Use of Interpretability Tools for Machine Learning}. In
  \bibinfo{booktitle}{\emph{Proceedings of the 2020 CHI Conference on Human
  Factors in Computing Systems}}. \bibinfo{pages}{1--14}.
\newblock


\bibitem[Kayacik et~al\mbox{.}(2019)]%
        {kayacik2019identifying}
\bibfield{author}{\bibinfo{person}{Claire Kayacik}, \bibinfo{person}{Sherol
  Chen}, \bibinfo{person}{Signe Noerly}, \bibinfo{person}{Jess Holbrook},
  \bibinfo{person}{Adam Roberts}, {and} \bibinfo{person}{Douglas Eck}.}
  \bibinfo{year}{2019}\natexlab{}.
\newblock \showarticletitle{Identifying the intersections: User experience+
  research scientist collaboration in a generative machine learning interface}.
  In \bibinfo{booktitle}{\emph{Extended Abstracts of the 2019 CHI Conference on
  Human Factors in Computing Systems}}. \bibinfo{pages}{1--8}.
\newblock


\bibitem[Keil(2006)]%
        {keil2006explanation}
\bibfield{author}{\bibinfo{person}{Frank~C Keil}.}
  \bibinfo{year}{2006}\natexlab{}.
\newblock \showarticletitle{Explanation and understanding}.
\newblock \bibinfo{journal}{\emph{Annu. Rev. Psychol.}}  \bibinfo{volume}{57}
  (\bibinfo{year}{2006}), \bibinfo{pages}{227--254}.
\newblock


\bibitem[Koh et~al\mbox{.}(2022)]%
        {koh2022empirical}
\bibfield{author}{\bibinfo{person}{Huan~Yee Koh}, \bibinfo{person}{Jiaxin Ju},
  \bibinfo{person}{Ming Liu}, {and} \bibinfo{person}{Shirui Pan}.}
  \bibinfo{year}{2022}\natexlab{}.
\newblock \showarticletitle{An Empirical Survey on Long Document Summarization:
  Datasets, Models and Metrics}.
\newblock \bibinfo{journal}{\emph{ACM Journal of the ACM (JACM)}}
  (\bibinfo{year}{2022}).
\newblock


\bibitem[Lee et~al\mbox{.}(2022)]%
        {lee2022coauthor}
\bibfield{author}{\bibinfo{person}{Mina Lee}, \bibinfo{person}{Percy Liang},
  {and} \bibinfo{person}{Qian Yang}.} \bibinfo{year}{2022}\natexlab{}.
\newblock \showarticletitle{Coauthor: Designing a human-ai collaborative
  writing dataset for exploring language model capabilities}. In
  \bibinfo{booktitle}{\emph{CHI Conference on Human Factors in Computing
  Systems}}. \bibinfo{pages}{1--19}.
\newblock


\bibitem[Liao et~al\mbox{.}(2020)]%
        {liao2020questioning}
\bibfield{author}{\bibinfo{person}{Q~Vera Liao}, \bibinfo{person}{Daniel
  Gruen}, {and} \bibinfo{person}{Sarah Miller}.}
  \bibinfo{year}{2020}\natexlab{}.
\newblock \showarticletitle{Questioning the AI: informing design practices for
  explainable AI user experiences}. In \bibinfo{booktitle}{\emph{Proceedings of
  the 2020 CHI Conference on Human Factors in Computing Systems}}.
  \bibinfo{pages}{1--15}.
\newblock


\bibitem[Liao et~al\mbox{.}(2021)]%
        {liao2021question}
\bibfield{author}{\bibinfo{person}{Q~Vera Liao}, \bibinfo{person}{Milena
  Pribi{\'c}}, \bibinfo{person}{Jaesik Han}, \bibinfo{person}{Sarah Miller},
  {and} \bibinfo{person}{Daby Sow}.} \bibinfo{year}{2021}\natexlab{}.
\newblock \showarticletitle{Question-driven design process for explainable ai
  user experiences}.
\newblock \bibinfo{journal}{\emph{arXiv preprint arXiv:2104.03483}}
  (\bibinfo{year}{2021}).
\newblock


\bibitem[Liao and Varshney(2021)]%
        {liao2021human}
\bibfield{author}{\bibinfo{person}{Q~Vera Liao} {and} \bibinfo{person}{Kush~R
  Varshney}.} \bibinfo{year}{2021}\natexlab{}.
\newblock \showarticletitle{Human-Centered Explainable AI (XAI): From
  Algorithms to User Experiences}.
\newblock \bibinfo{journal}{\emph{arXiv preprint arXiv:2110.10790}}
  (\bibinfo{year}{2021}).
\newblock


\bibitem[Liao et~al\mbox{.}(2022)]%
        {liao2022connecting}
\bibfield{author}{\bibinfo{person}{Q~Vera Liao}, \bibinfo{person}{Yunfeng
  Zhang}, \bibinfo{person}{Ronny Luss}, \bibinfo{person}{Finale Doshi-Velez},
  {and} \bibinfo{person}{Amit Dhurandhar}.} \bibinfo{year}{2022}\natexlab{}.
\newblock \showarticletitle{Connecting Algorithmic Research and Usage Contexts:
  A Perspective of Contextualized Evaluation for Explainable AI}. In
  \bibinfo{booktitle}{\emph{Proceedings of the AAAI Conference on Human
  Computation and Crowdsourcing}}, Vol.~\bibinfo{volume}{10}.
  \bibinfo{pages}{147--159}.
\newblock


\bibitem[Lipton(2018)]%
        {lipton2018mythos}
\bibfield{author}{\bibinfo{person}{Zachary~C Lipton}.}
  \bibinfo{year}{2018}\natexlab{}.
\newblock \showarticletitle{The Mythos of Model Interpretability: In machine
  learning, the concept of interpretability is both important and slippery.}
\newblock \bibinfo{journal}{\emph{Queue}} \bibinfo{volume}{16},
  \bibinfo{number}{3} (\bibinfo{year}{2018}), \bibinfo{pages}{31--57}.
\newblock


\bibitem[Lombrozo(2006)]%
        {lombrozo2006structure}
\bibfield{author}{\bibinfo{person}{Tania Lombrozo}.}
  \bibinfo{year}{2006}\natexlab{}.
\newblock \showarticletitle{The structure and function of explanations}.
\newblock \bibinfo{journal}{\emph{Trends in cognitive sciences}}
  \bibinfo{volume}{10}, \bibinfo{number}{10} (\bibinfo{year}{2006}),
  \bibinfo{pages}{464--470}.
\newblock


\bibitem[Lombrozo(2012)]%
        {lombrozo2012explanation}
\bibfield{author}{\bibinfo{person}{Tania Lombrozo}.}
  \bibinfo{year}{2012}\natexlab{}.
\newblock \showarticletitle{Explanation and abductive inference.}
\newblock  (\bibinfo{year}{2012}).
\newblock


\bibitem[Lu et~al\mbox{.}(2021)]%
        {lu2021impact}
\bibfield{author}{\bibinfo{person}{Jiahao Lu}, \bibinfo{person}{Alejandra~Gomez
  Ortega}, \bibinfo{person}{Milene Gon{\c{c}}alves}, {and}
  \bibinfo{person}{Jacky Bourgeois}.} \bibinfo{year}{2021}\natexlab{}.
\newblock \showarticletitle{The Impact of Data on the Role of Designers and
  their Process}.
\newblock \bibinfo{journal}{\emph{Proceedings of the Design Society}}
  \bibinfo{volume}{1} (\bibinfo{year}{2021}), \bibinfo{pages}{3021--3030}.
\newblock


\bibitem[Madaio et~al\mbox{.}(2022)]%
        {madaio2022assessing}
\bibfield{author}{\bibinfo{person}{Michael Madaio}, \bibinfo{person}{Lisa
  Egede}, \bibinfo{person}{Hariharan Subramonyam}, \bibinfo{person}{Jennifer
  Wortman~Vaughan}, {and} \bibinfo{person}{Hanna Wallach}.}
  \bibinfo{year}{2022}\natexlab{}.
\newblock \showarticletitle{Assessing the Fairness of AI Systems: AI
  Practitioners' Processes, Challenges, and Needs for Support}.
\newblock \bibinfo{journal}{\emph{Proceedings of the ACM on Human-Computer
  Interaction}} \bibinfo{volume}{6}, \bibinfo{number}{CSCW1}
  (\bibinfo{year}{2022}).
\newblock


\bibitem[Madaio et~al\mbox{.}(2020)]%
        {madaio2020co}
\bibfield{author}{\bibinfo{person}{Michael~A Madaio}, \bibinfo{person}{Luke
  Stark}, \bibinfo{person}{Jennifer Wortman~Vaughan}, {and}
  \bibinfo{person}{Hanna Wallach}.} \bibinfo{year}{2020}\natexlab{}.
\newblock \showarticletitle{Co-designing checklists to understand
  organizational challenges and opportunities around fairness in AI}. In
  \bibinfo{booktitle}{\emph{Proceedings of the 2020 CHI Conference on Human
  Factors in Computing Systems}}. \bibinfo{pages}{1--14}.
\newblock


\bibitem[Malsattar et~al\mbox{.}(2019)]%
        {malsattar2019designing}
\bibfield{author}{\bibinfo{person}{Nirav Malsattar}, \bibinfo{person}{Tomo
  Kihara}, {and} \bibinfo{person}{Elisa Giaccardi}.}
  \bibinfo{year}{2019}\natexlab{}.
\newblock \showarticletitle{Designing and Prototyping from the Perspective of
  AI in the Wild}. In \bibinfo{booktitle}{\emph{Proceedings of the 2019 on
  Designing Interactive Systems Conference}}. \bibinfo{pages}{1083--1088}.
\newblock


\bibitem[Miller(2019)]%
        {miller2019explanation}
\bibfield{author}{\bibinfo{person}{Tim Miller}.}
  \bibinfo{year}{2019}\natexlab{}.
\newblock \showarticletitle{Explanation in artificial intelligence: Insights
  from the social sciences}.
\newblock \bibinfo{journal}{\emph{Artificial intelligence}}
  \bibinfo{volume}{267} (\bibinfo{year}{2019}), \bibinfo{pages}{1--38}.
\newblock


\bibitem[Mitchell et~al\mbox{.}(2019)]%
        {mitchell2019model}
\bibfield{author}{\bibinfo{person}{Margaret Mitchell}, \bibinfo{person}{Simone
  Wu}, \bibinfo{person}{Andrew Zaldivar}, \bibinfo{person}{Parker Barnes},
  \bibinfo{person}{Lucy Vasserman}, \bibinfo{person}{Ben Hutchinson},
  \bibinfo{person}{Elena Spitzer}, \bibinfo{person}{Inioluwa~Deborah Raji},
  {and} \bibinfo{person}{Timnit Gebru}.} \bibinfo{year}{2019}\natexlab{}.
\newblock \showarticletitle{Model cards for model reporting}. In
  \bibinfo{booktitle}{\emph{Proceedings of the conference on fairness,
  accountability, and transparency}}. \bibinfo{pages}{220--229}.
\newblock


\bibitem[Park et~al\mbox{.}(2022)]%
        {park2022designing}
\bibfield{author}{\bibinfo{person}{Hyanghee Park}, \bibinfo{person}{Daehwan
  Ahn}, \bibinfo{person}{Kartik Hosanagar}, {and} \bibinfo{person}{Joonhwan
  Lee}.} \bibinfo{year}{2022}\natexlab{}.
\newblock \showarticletitle{Designing Fair AI in Human Resource Management:
  Understanding Tensions Surrounding Algorithmic Evaluation and Envisioning
  Stakeholder-Centered Solutions}. In \bibinfo{booktitle}{\emph{CHI Conference
  on Human Factors in Computing Systems}}. \bibinfo{pages}{1--22}.
\newblock


\bibitem[{Partnership on AI}(2021)]%
        {aboutml}
\bibfield{author}{\bibinfo{person}{{Partnership on AI}}.}
  \bibinfo{year}{2021}\natexlab{}.
\newblock \bibinfo{booktitle}{\emph{ABOUT ML Reference Document}}.
\newblock \bibinfo{type}{{T}echnical {R}eport}.
\newblock
\newblock
\shownote{https://partnershiponai.org/paper/about-ml-reference-document/}.


\bibitem[Raji et~al\mbox{.}(2020)]%
        {raji2020closing}
\bibfield{author}{\bibinfo{person}{Inioluwa~Deborah Raji},
  \bibinfo{person}{Andrew Smart}, \bibinfo{person}{Rebecca~N White},
  \bibinfo{person}{Margaret Mitchell}, \bibinfo{person}{Timnit Gebru},
  \bibinfo{person}{Ben Hutchinson}, \bibinfo{person}{Jamila Smith-Loud},
  \bibinfo{person}{Daniel Theron}, {and} \bibinfo{person}{Parker Barnes}.}
  \bibinfo{year}{2020}\natexlab{}.
\newblock \showarticletitle{Closing the AI accountability gap: Defining an
  end-to-end framework for internal algorithmic auditing}. In
  \bibinfo{booktitle}{\emph{Proceedings of the 2020 conference on fairness,
  accountability, and transparency}}. \bibinfo{pages}{33--44}.
\newblock


\bibitem[Rakova et~al\mbox{.}(2021)]%
        {rakova2021responsible}
\bibfield{author}{\bibinfo{person}{Bogdana Rakova}, \bibinfo{person}{Jingying
  Yang}, \bibinfo{person}{Henriette Cramer}, {and} \bibinfo{person}{Rumman
  Chowdhury}.} \bibinfo{year}{2021}\natexlab{}.
\newblock \showarticletitle{Where responsible AI meets reality: Practitioner
  perspectives on enablers for shifting organizational practices}.
\newblock \bibinfo{journal}{\emph{Proceedings of the ACM on Human-Computer
  Interaction}} \bibinfo{volume}{5}, \bibinfo{number}{CSCW1}
  (\bibinfo{year}{2021}), \bibinfo{pages}{1--23}.
\newblock


\bibitem[Ramesh et~al\mbox{.}(2021)]%
        {ramesh2021zero}
\bibfield{author}{\bibinfo{person}{Aditya Ramesh}, \bibinfo{person}{Mikhail
  Pavlov}, \bibinfo{person}{Gabriel Goh}, \bibinfo{person}{Scott Gray},
  \bibinfo{person}{Chelsea Voss}, \bibinfo{person}{Alec Radford},
  \bibinfo{person}{Mark Chen}, {and} \bibinfo{person}{Ilya Sutskever}.}
  \bibinfo{year}{2021}\natexlab{}.
\newblock \showarticletitle{Zero-shot text-to-image generation}. In
  \bibinfo{booktitle}{\emph{International Conference on Machine Learning}}.
  PMLR, \bibinfo{pages}{8821--8831}.
\newblock


\bibitem[Ribeiro et~al\mbox{.}(2020)]%
        {ribeiro2020beyond}
\bibfield{author}{\bibinfo{person}{Marco~Tulio Ribeiro},
  \bibinfo{person}{Tongshuang Wu}, \bibinfo{person}{Carlos Guestrin}, {and}
  \bibinfo{person}{Sameer Singh}.} \bibinfo{year}{2020}\natexlab{}.
\newblock \showarticletitle{Beyond Accuracy: Behavioral Testing of NLP Models
  with CheckList}. In \bibinfo{booktitle}{\emph{Proceedings of the 58th Annual
  Meeting of the Association for Computational Linguistics}}.
  \bibinfo{pages}{4902--4912}.
\newblock


\bibitem[Robles and Wiberg(2010)]%
        {robles2010texturing}
\bibfield{author}{\bibinfo{person}{Erica Robles} {and} \bibinfo{person}{Mikael
  Wiberg}.} \bibinfo{year}{2010}\natexlab{}.
\newblock \showarticletitle{Texturing the" material turn" in interaction
  design}. In \bibinfo{booktitle}{\emph{Proceedings of the fourth international
  conference on Tangible, embedded, and embodied interaction}}.
  \bibinfo{pages}{137--144}.
\newblock


\bibitem[Savolainen(1993)]%
        {savolainen1993sense}
\bibfield{author}{\bibinfo{person}{Reijo Savolainen}.}
  \bibinfo{year}{1993}\natexlab{}.
\newblock \showarticletitle{The sense-making theory: Reviewing the interests of
  a user-centered approach to information seeking and use}.
\newblock \bibinfo{journal}{\emph{Information processing \& management}}
  \bibinfo{volume}{29}, \bibinfo{number}{1} (\bibinfo{year}{1993}),
  \bibinfo{pages}{13--28}.
\newblock


\bibitem[Schon and Bennett(1996)]%
        {schon1996reflective}
\bibfield{author}{\bibinfo{person}{Donald Schon} {and} \bibinfo{person}{John
  Bennett}.} \bibinfo{year}{1996}\natexlab{}.
\newblock \bibinfo{title}{Reflective Conversation with Materials in Bringing
  Design to Software, Winograd T}.
\newblock
\newblock


\bibitem[Shneiderman(2021)]%
        {shneiderman2021responsible}
\bibfield{author}{\bibinfo{person}{Ben Shneiderman}.}
  \bibinfo{year}{2021}\natexlab{}.
\newblock \showarticletitle{Responsible AI: Bridging from ethics to practice}.
\newblock \bibinfo{journal}{\emph{Commun. ACM}} \bibinfo{volume}{64},
  \bibinfo{number}{8} (\bibinfo{year}{2021}), \bibinfo{pages}{32--35}.
\newblock


\bibitem[Sloane et~al\mbox{.}(2020)]%
        {sloane2020participation}
\bibfield{author}{\bibinfo{person}{Mona Sloane}, \bibinfo{person}{Emanuel
  Moss}, \bibinfo{person}{Olaitan Awomolo}, {and} \bibinfo{person}{Laura
  Forlano}.} \bibinfo{year}{2020}\natexlab{}.
\newblock \showarticletitle{Participation is not a design fix for machine
  learning}.
\newblock \bibinfo{journal}{\emph{arXiv preprint arXiv:2007.02423}}
  (\bibinfo{year}{2020}).
\newblock


\bibitem[Subramonyam et~al\mbox{.}(2022)]%
        {subramonyam2022solving}
\bibfield{author}{\bibinfo{person}{Hariharan Subramonyam},
  \bibinfo{person}{Jane Im}, \bibinfo{person}{Colleen Seifert}, {and}
  \bibinfo{person}{Eytan Adar}.} \bibinfo{year}{2022}\natexlab{}.
\newblock \showarticletitle{Solving Separation-of-Concerns Problems in
  Collaborative Design of Human-AI Systems through Leaky Abstractions}. In
  \bibinfo{booktitle}{\emph{CHI Conference on Human Factors in Computing
  Systems}}. \bibinfo{pages}{1--21}.
\newblock


\bibitem[Subramonyam et~al\mbox{.}(2021a)]%
        {subramonyam2021protoai}
\bibfield{author}{\bibinfo{person}{Hariharan Subramonyam},
  \bibinfo{person}{Colleen Seifert}, {and} \bibinfo{person}{Eytan Adar}.}
  \bibinfo{year}{2021}\natexlab{a}.
\newblock \showarticletitle{ProtoAI: Model-Informed Prototyping for AI-Powered
  Interfaces}. In \bibinfo{booktitle}{\emph{26th International Conference on
  Intelligent User Interfaces}}. \bibinfo{pages}{48--58}.
\newblock


\bibitem[Subramonyam et~al\mbox{.}(2021b)]%
        {subramonyam2021towards}
\bibfield{author}{\bibinfo{person}{Hariharan Subramonyam},
  \bibinfo{person}{Colleen Seifert}, {and} \bibinfo{person}{Eytan Adar}.}
  \bibinfo{year}{2021}\natexlab{b}.
\newblock \showarticletitle{Towards a process model for co-creating AI
  experiences}. In \bibinfo{booktitle}{\emph{Designing Interactive Systems
  Conference 2021}}. \bibinfo{pages}{1529--1543}.
\newblock


\bibitem[Suresh et~al\mbox{.}(2021)]%
        {suresh2021beyond}
\bibfield{author}{\bibinfo{person}{Harini Suresh}, \bibinfo{person}{Steven~R
  Gomez}, \bibinfo{person}{Kevin~K Nam}, {and} \bibinfo{person}{Arvind
  Satyanarayan}.} \bibinfo{year}{2021}\natexlab{}.
\newblock \showarticletitle{Beyond expertise and roles: A framework to
  characterize the stakeholders of interpretable machine learning and their
  needs}. In \bibinfo{booktitle}{\emph{Proceedings of the 2021 CHI Conference
  on Human Factors in Computing Systems}}. \bibinfo{pages}{1--16}.
\newblock


\bibitem[Weidinger et~al\mbox{.}(2022)]%
        {weidinger2022taxonomy}
\bibfield{author}{\bibinfo{person}{Laura Weidinger}, \bibinfo{person}{Jonathan
  Uesato}, \bibinfo{person}{Maribeth Rauh}, \bibinfo{person}{Conor Griffin},
  \bibinfo{person}{Po-Sen Huang}, \bibinfo{person}{John Mellor},
  \bibinfo{person}{Amelia Glaese}, \bibinfo{person}{Myra Cheng},
  \bibinfo{person}{Borja Balle}, \bibinfo{person}{Atoosa Kasirzadeh},
  {et~al\mbox{.}}} \bibinfo{year}{2022}\natexlab{}.
\newblock \showarticletitle{Taxonomy of risks posed by language models}. In
  \bibinfo{booktitle}{\emph{2022 ACM Conference on Fairness, Accountability,
  and Transparency}}. \bibinfo{pages}{214--229}.
\newblock


\bibitem[Windl et~al\mbox{.}(2022)]%
        {windl2022not}
\bibfield{author}{\bibinfo{person}{Maximiliane Windl},
  \bibinfo{person}{Sebastian~S Feger}, \bibinfo{person}{Lara Zijlstra},
  \bibinfo{person}{Albrecht Schmidt}, {and} \bibinfo{person}{Pawel~W Wozniak}.}
  \bibinfo{year}{2022}\natexlab{}.
\newblock \showarticletitle{‘It Is Not Always Discovery Time’: Four
  Pragmatic Approaches in Designing AI Systems}. In
  \bibinfo{booktitle}{\emph{CHI Conference on Human Factors in Computing
  Systems}}. \bibinfo{pages}{1--12}.
\newblock


\bibitem[Wolf et~al\mbox{.}(2019)]%
        {wolf2019huggingface}
\bibfield{author}{\bibinfo{person}{Thomas Wolf}, \bibinfo{person}{Lysandre
  Debut}, \bibinfo{person}{Victor Sanh}, \bibinfo{person}{Julien Chaumond},
  \bibinfo{person}{Clement Delangue}, \bibinfo{person}{Anthony Moi},
  \bibinfo{person}{Pierric Cistac}, \bibinfo{person}{Tim Rault},
  \bibinfo{person}{R{\'e}mi Louf}, \bibinfo{person}{Morgan Funtowicz},
  {et~al\mbox{.}}} \bibinfo{year}{2019}\natexlab{}.
\newblock \showarticletitle{Huggingface's transformers: State-of-the-art
  natural language processing}.
\newblock \bibinfo{journal}{\emph{arXiv preprint arXiv:1910.03771}}
  (\bibinfo{year}{2019}).
\newblock


\bibitem[Wortman~Vaughan and Wallach(2021)]%
        {vaughan2021humancentered}
\bibfield{author}{\bibinfo{person}{Jennifer Wortman~Vaughan} {and}
  \bibinfo{person}{Hanna Wallach}.} \bibinfo{year}{2021}\natexlab{}.
\newblock \showarticletitle{A Human-Centered Agenda for Intelligible Machine
  Learning}.
\newblock In \bibinfo{booktitle}{\emph{Machines We Trust: Perspectives on
  Dependable AI}}, \bibfield{editor}{\bibinfo{person}{Marcello Pelillo} {and}
  \bibinfo{person}{Teresa Scantamburlo}} (Eds.). \bibinfo{publisher}{MIT
  Press}.
\newblock


\bibitem[Wu et~al\mbox{.}(2019)]%
        {wu2019errudite}
\bibfield{author}{\bibinfo{person}{Tongshuang Wu}, \bibinfo{person}{Marco~Tulio
  Ribeiro}, \bibinfo{person}{Jeffrey Heer}, {and} \bibinfo{person}{Daniel~S
  Weld}.} \bibinfo{year}{2019}\natexlab{}.
\newblock \showarticletitle{Errudite: Scalable, reproducible, and testable
  error analysis}. In \bibinfo{booktitle}{\emph{Proceedings of the 57th Annual
  Meeting of the Association for Computational Linguistics}}.
  \bibinfo{pages}{747--763}.
\newblock


\bibitem[Yadav et~al\mbox{.}(2022)]%
        {yadav2022automatic}
\bibfield{author}{\bibinfo{person}{Divakar Yadav}, \bibinfo{person}{Jalpa
  Desai}, {and} \bibinfo{person}{Arun~Kumar Yadav}.}
  \bibinfo{year}{2022}\natexlab{}.
\newblock \showarticletitle{Automatic Text Summarization Methods: A
  Comprehensive Review}.
\newblock \bibinfo{journal}{\emph{arXiv preprint arXiv:2204.01849}}
  (\bibinfo{year}{2022}).
\newblock


\bibitem[Yang(2018)]%
        {yang2018machine}
\bibfield{author}{\bibinfo{person}{Qian Yang}.}
  \bibinfo{year}{2018}\natexlab{}.
\newblock \showarticletitle{Machine learning as a UX design material: how can
  we imagine beyond automation, recommenders, and reminders?}. In
  \bibinfo{booktitle}{\emph{AAAI Spring Symposia}}.
\newblock


\bibitem[Yang et~al\mbox{.}(2019)]%
        {yang2019sketching}
\bibfield{author}{\bibinfo{person}{Qian Yang}, \bibinfo{person}{Justin
  Cranshaw}, \bibinfo{person}{Saleema Amershi}, \bibinfo{person}{Shamsi~T
  Iqbal}, {and} \bibinfo{person}{Jaime Teevan}.}
  \bibinfo{year}{2019}\natexlab{}.
\newblock \showarticletitle{Sketching nlp: A case study of exploring the right
  things to design with language intelligence}. In
  \bibinfo{booktitle}{\emph{Proceedings of the 2019 CHI Conference on Human
  Factors in Computing Systems}}. \bibinfo{pages}{1--12}.
\newblock


\bibitem[Yang et~al\mbox{.}(2018)]%
        {yang2018investigating}
\bibfield{author}{\bibinfo{person}{Qian Yang}, \bibinfo{person}{Alex Scuito},
  \bibinfo{person}{John Zimmerman}, \bibinfo{person}{Jodi Forlizzi}, {and}
  \bibinfo{person}{Aaron Steinfeld}.} \bibinfo{year}{2018}\natexlab{}.
\newblock \showarticletitle{Investigating how experienced UX designers
  effectively work with machine learning}. In
  \bibinfo{booktitle}{\emph{Proceedings of the 2018 designing interactive
  systems conference}}. \bibinfo{pages}{585--596}.
\newblock


\bibitem[Yang et~al\mbox{.}(2020)]%
        {yang2020re}
\bibfield{author}{\bibinfo{person}{Qian Yang}, \bibinfo{person}{Aaron
  Steinfeld}, \bibinfo{person}{Carolyn Ros{\'e}}, {and} \bibinfo{person}{John
  Zimmerman}.} \bibinfo{year}{2020}\natexlab{}.
\newblock \showarticletitle{Re-examining whether, why, and how human-AI
  interaction is uniquely difficult to design}. In
  \bibinfo{booktitle}{\emph{Proceedings of the 2020 chi conference on human
  factors in computing systems}}. \bibinfo{pages}{1--13}.
\newblock


\bibitem[Yildirim et~al\mbox{.}(2022)]%
        {yildirim2022experienced}
\bibfield{author}{\bibinfo{person}{Nur Yildirim}, \bibinfo{person}{Alex Kass},
  \bibinfo{person}{Teresa Tung}, \bibinfo{person}{Connor Upton},
  \bibinfo{person}{Donnacha Costello}, \bibinfo{person}{Robert Giusti},
  \bibinfo{person}{Sinem Lacin}, \bibinfo{person}{Sara Lovic},
  \bibinfo{person}{James~M O'Neill}, \bibinfo{person}{Rudi~O'Reilly Meehan},
  {et~al\mbox{.}}} \bibinfo{year}{2022}\natexlab{}.
\newblock \showarticletitle{How Experienced Designers of Enterprise
  Applications Engage AI as a Design Material}. In
  \bibinfo{booktitle}{\emph{CHI Conference on Human Factors in Computing
  Systems}}. \bibinfo{pages}{1--13}.
\newblock


\bibitem[Yu et~al\mbox{.}(2020)]%
        {yu2020keeping}
\bibfield{author}{\bibinfo{person}{Bowen Yu}, \bibinfo{person}{Ye Yuan},
  \bibinfo{person}{Loren Terveen}, \bibinfo{person}{Zhiwei~Steven Wu},
  \bibinfo{person}{Jodi Forlizzi}, {and} \bibinfo{person}{Haiyi Zhu}.}
  \bibinfo{year}{2020}\natexlab{}.
\newblock \showarticletitle{Keeping designers in the loop: Communicating
  inherent algorithmic trade-offs across multiple objectives}. In
  \bibinfo{booktitle}{\emph{Proceedings of the 2020 ACM designing interactive
  systems conference}}. \bibinfo{pages}{1245--1257}.
\newblock


\bibitem[Zdanowska and Taylor(2022)]%
        {zdanowska2022study}
\bibfield{author}{\bibinfo{person}{Sabah Zdanowska} {and}
  \bibinfo{person}{Alex~S Taylor}.} \bibinfo{year}{2022}\natexlab{}.
\newblock \showarticletitle{A study of UX practitioners roles in designing
  real-world, enterprise ML systems}. In \bibinfo{booktitle}{\emph{CHI
  Conference on Human Factors in Computing Systems}}. \bibinfo{pages}{1--15}.
\newblock


\end{thebibliography}

\onecolumn
\newpage

\section{Appendix}

\begin{table*}[!h]
\small
\sffamily
    \centering
    \begin{tabular}{cc}
     \textbf{\rv{Category}}    & \textbf{\rv{Content}}  \\
      \toprule
      
      \begin{tabular}{>{\centering\arraybackslash}p{2cm}}
      \rv{Model description} 
       \end{tabular}
       &\begin{tabular}{p{10.5cm}}\rv{The extractive summarization model uses natural language processing techniques to locate key sentences in an unstructured text document. These sentences collectively convey the main idea of the document. }
         
     \rv{When a document is given as the input, the model returns a list of extracted sentences, together with a rank score and its position in the original document for each extracted sentence. A rank score is an indicator of how relevant or important the model considers the sentence is to the main idea of the document (between 0 and 1, higher as more relevant). }
         
    \rv{By default, the model returns three highest scored sentences, and you can specify the number of sentences returned.}
         \end{tabular}
         \\
         \midrule
         
    \begin{tabular}{>{\centering\arraybackslash}p{2cm}}
      \rv{Examples of intended use}
       \end{tabular}
       &\begin{tabular}{p{10.5cm}}
       
    \rv{You might want to use the extractive summarization model to:}
    \begin{itemize}
        \item \rv{Distill critical information from lengthy documents.}
        \item \rv{Highlight key sentences in documents.}
        \item \rv{Quickly skim documents in a library.}
        \item \rv{Generate news feed content.}
    \end{itemize}{}
\rv{You can use extractive summarization in multiple scenarios across a variety of industries. For example:}
\begin{itemize}
    \item \rv{Extract key information from public news articles to produce insights.}
    \item \rv{Classify documents by their key contents.}
    \item \rv{Distill important information from long documents to empower solutions such as search, question and answering, and decision support.}
    \vspace{-0.2cm}
\end{itemize}{}

       \end{tabular}
\\
\midrule    
         
   \begin{tabular}{>{\centering\arraybackslash}p{2cm}}
        \rv{Do not use (Unintended use)}
       \end{tabular}
       &\begin{tabular}{p{10.5cm}}
       \rv{Don't use extractive summarization for automatic actions without human intervention for high-impact scenarios. A person should always review source data when another person's economic situation, health, or safety is affected.}
       \end{tabular}
    \\    

    \midrule
    
       \begin{tabular}{>{\centering\arraybackslash}p{2cm}}
      \rv{Limitations with impacting factors}
       \end{tabular}
       &
       \begin{tabular}{p{10.5cm}}
       \rv{Based on your scenario and input data, you could experience different levels of performance:}
       \begin{itemize}
           \item \rv{Because the model is trained on document-based texts, such as news articles, scientific reports, and legal documents, when used with texts in certain genres such as conversations and transcriptions, it might produce output with lower accuracy.}
           \item \rv{When used with texts that may contain errors or are less similar to well-formed sentences, such as texts extracted from lists, tables, charts, the model might produce output with lower accuracy.}
             \vspace{-0.2cm}
       \end{itemize}{}
       \end{tabular}
       \\
       \midrule    
         
   \begin{tabular}{>{\centering\arraybackslash}p{2cm}}
       \rv{Design Space Guidance}
       \end{tabular}
       &\begin{tabular}{p{10.5cm}}
        \rv{\textbf{Input}: How to align inputs to what work best for AI?}
        
          \rv{ \textbf{Output}: How to present AI outputs to users?}
           
          \rv{ \textbf{Failure}: How to handle AI errors and provide paths from failure?}
           
          \rv{  \textbf{Transparency}: How to support user understanding of AI and AI outputs?}
            
         \rv{ \textbf{Feedback}: How to support users providing feedback for AI to learn?}
       \end{tabular}
    \\    
 \midrule    
         
   \begin{tabular}{>{\centering\arraybackslash}p{2cm}}
       \rv{ Harms considerations}
       \end{tabular}
       &\begin{tabular}{p{10.5cm}}
\rv{Here is a description of general \textbf{technical limitations} and \textbf{\textit{potential harms}} for summarization models. }

\rv{\textbf{Performance biases}: it may work less well on articles that are less structured, contain informal language, longer, or on topics that were less common in the training data. This could lead to \textbf{\textit{disparate impacts}} for users reading different topics, sources, language styles, etc.}
 
 \rv{\textbf{Structural biases}: it may be biased towards extracting from the beginning part of an article or paragraphs. This could lead to the \textbf{\textit{erasure of perspectives}} or \textbf{\textit{misinformation}}.}
   
  \rv{ \textbf{Limits in extraction and linguistic quality}: it may fail to extract sentences with words that are out of the model's vocabulary. The extracted sentences may be incomplete or repetitive. This could lead to \textbf{\textit{misinformation}}, \textbf{\textit{erasure of perspectives}}, and low-quality even \textbf{\textit{offensive content}} to the audience}
       \end{tabular}
    \\    
 \bottomrule

           \end{tabular}
    \caption{\rv{Content of model documentation presented to participants. Original images are presented in Section~\ref{sec:method}.}}
    \label{content}
\end{table*}{}

\begin{table*}[]
\small
    \centering
    \begin{tabular}{cccccc}
       \textbf{ID}  & \textbf{Role} & \textbf{Years in profession}& \textbf{Experience with AI design} & \textbf{Experience with NLP}& \textbf{\rv{Gender}}  \\
       \toprule
       1 & UX researcher & 1--5 years & Part of my day-to-day job& Part of my day-to-day job& \rv{Female}
       \\
       \midrule
       
       2 & HCI researcher & 1--5 years & Part of my day-to-day job& Never& \rv{Female}
              \\
       \midrule
       
       3 & Product Designer & 5--10 years & Part of my day-to-day job& Part of my day-to-day job& \rv{Male}
       
    \\
       \midrule
       
       4 & Product Designer & 5--10 years & Part of my day-to-day job& Part of my day-to-day job& \rv{Female}
               \\
       \midrule
       
       5 & Interaction designer & 1--5 years & Part of my day-to-day job& Limited experience& \rv{Female}
    
                   \\
       \midrule
       
       6 & UI/UX designer & 1--5 years & Part of my day-to-day job& Limited experience& \rv{Female}
       
                          \\
       \midrule
       
       7 & Designer & 5--10 years & Part of my day-to-day job& Limited experience& \rv{Female}
                                 \\
       \midrule
       
       8 & Product designer & 5--10 years & Limited experience & Never& \rv{Female}
       
                                        \\
       \midrule
       
       9 & Design lead & More than 10 years & I consider myself an expert & Part of my day-to-day job& \rv{Male}
       \\
       \midrule
       
       10 & Product Manager & 1--5 years& Part of my day-to-day job & Limited experience& \rv{Male}
       
              \\
       \midrule
       
       11 & Product designer & 5--10 years& Limited experience
        & Never& \rv{Male}
        
                      \\
       \midrule
       
       12 & Interaction designer & 1--5 years& Part of my day-to-day job
        & Part of my day-to-day job& \rv{Female}
        
                              \\
       \midrule
       
       13 & Designer &More than 10 years& Part of my day-to-day job
        & Part of my day-to-day job& \rv{Male}
        
                                      \\
       \midrule
       
       14 &User Researcher &1--5 years& Part of my day-to-day job
        & Part of my day-to-day job& \rv{Female}
       
                                             \\
       \midrule
       
       15 &Product designer  &More than 10 years& Limited experience
        & Limited experience& \rv{Female}
        
                                                     \\
       \midrule
       
       16 &UX researcher  &5--10 years& Part of my day-to-day job
        & Limited experience& \rv{Female}
        
                                                   \\
       \midrule
       
       17 &Designer  &5--10 years& Limited experience
        & Limited experience& \rv{Male}
        
                                                           \\
       \midrule
       
       18 &Interaction designer  &1--5 years& Limited experience
        & Limited experience& \rv{Female}
        
                                                                  \\
       \midrule
       
       19 &Product Designer  &1--5 years& Part of my day-to-day job
        & Limited experience& \rv{Female}
       
        \\
       \midrule
       
       20 &UX Designer  &1--5 years& Limited experience
        & Never& \rv{Female}
              
        \\
       \midrule
       
       21 &UX Designer  &5--10 years& Part of my day-to-day job
        & Part of my day-to-day job& \rv{Male}
        
                \\
       \midrule
       
       22 &Product manager  &1--5 years& I consider myself an expert
        & Never& \rv{Male}
        
                       \\
       \midrule
       
       22 &Product manager  &More than 10 years&Part of my day-to-day& I consider myself an expert& \rv{Female}
       \\
       \bottomrule
       
    \end{tabular}
    \caption{Description of participants.}
    \label{demographics}
\end{table*}{}

\begin{table*}[h]
\small{
\sffamily
    \centering
    \begin{tabular}{ccc}
   
    \textbf{Information Category} & \textbf{Summary} & \textbf{Example Quotes}\\
     \toprule
    \multicolumn{3}{c}{Provided in the Task}\\
    \toprule
    \begin{tabular}{>{\centering\arraybackslash}p{1.4cm}}Harms considerations \\(N=13)\end{tabular}&
    \begin{tabular}{p{6cm}}Participants appreciated the awareness of potential harms, and the delineation of different sources of technical limitations.
    
    \vspace{0.1cm}
    However, many struggled with not having a concrete understanding and a lack of actionability to address these harms.
    \end{tabular}
        &  
    \begin{tabular}{p{5cm}}
    \textit{This was really useful. I realized I was talking about performance biases and structural biases in the same way... that helped me think more granularly, how I should design to address each of these sources} (P2)
    
     \vspace{0.1cm}
  \textit{The potential harms... those were a little abstract...I have a hard time thinking about actual instances} (P4)
    \end{tabular}
         
         \\
         
 \midrule

    \begin{tabular}{>{\centering\arraybackslash}p{1.4cm}} Impacting factors  \\(N=12)\end{tabular}&
    \begin{tabular}{p{6cm}}
    Some picked up the factor of article structure mentioned in this section and considered not applying the model to unstructured inputs.
    
     \vspace{0.1cm}
    However, the majority expressed dissatisfaction because they wished to know whether other factors, such as article length, genre, and language style, can impact the model.
    
     \vspace{0.1cm}
    Also wished to understand how the model behaves differently (e.g., output length, \rv{frequency of certain words}), rather than just how the performance varies, with factors.
    \end{tabular}
        &  
    \begin{tabular}{p{5cm}}
    \textit{I made a lot of assumptions that could be more well informed. Like is it better for things that are factual or opinion pieces? } (P12)

 \vspace{0.1cm}
    \textit{Will there be concept that is harder for the model [to extract]...  what exactly are good for providing these kinds of output is not clear to me (P21)
    }
    \end{tabular}
         
         \\
         
 \midrule

    \begin{tabular}{>{\centering\arraybackslash}p{1.4cm}} Examples from playground \\(N=11)\end{tabular}&
    \begin{tabular}{p{6cm}}
   Appreciated that the output example provided an intuitive understanding of the model affordance.
   
    \vspace{0.1cm}

    Experienced designers were intentional in examining different types of input-output pairs and looked for edge cases, often to explore the reliability of the model, and to discover or examine the effect of impacting factors.
    \end{tabular}
        &  
    \begin{tabular}{p{5cm}}
\textit{It is helpful to understand different ways the model could be used, like you don't have to just use the output, you can also rank the sentences, you can use the sentences within the context of the article} (P2)

\vspace{0.1cm}

\textit{I picked the statement of Ukraine because I'm assuming it talks about sensitive matters... Because I know language models are problematic when it comes to sensitive issues} (P3)
    \end{tabular}
         
         \\
         
 \midrule
 
    \begin{tabular}{>{\centering\arraybackslash}p{1.4cm}} Design space guidance \\(N=8)\end{tabular}&
    \begin{tabular}{p{6cm}}
Appreciated it as a checklist to help them think systematically about what to design for, especially for those new to AI design.

 \vspace{0.1cm}

As both a generative tool for inspiring designs and an evaluative tool to ensure the design quality.

 \vspace{0.1cm}

Helpful for setting common languages and goals when communicating with other team members.  
    \end{tabular}
        &  
    \begin{tabular}{p{5cm}}
\textit{The first is, as a generative tool...provides you a thing to think about, to apply to the designs that are in progress... Secondarily, it can provide a checklist for quality assurance.} (P9)

 \vspace{0.1cm}
\textit{Designing with and for AI is a relatively emerging territory. So just even being able to flag that in a way that's shared across the team would be really useful.} (P7)
    \end{tabular}
         
         \\ 
 \midrule

    \begin{tabular}{>{\centering\arraybackslash}p{1.4cm}} Do not use \\(N=7)\end{tabular}&
    \begin{tabular}{p{6cm}}
Appreciated documentation that leads with critical information. Some picked up the mentioning of avoiding use in ``high-stakes'' situations in their design thinking.

 \vspace{0.1cm}

However, the section was too high-level. Needed more examples of out-of-scope scenarios and understanding of outcomes.
    \end{tabular}
        &  
    \begin{tabular}{p{5cm}}
\textit{I personally look at AI from a very critical lens. So I naturally gravitate towards things that talk about limitations and do not use.} (P1)

 \vspace{0.1cm}

\textit{I'm a little confused on high impact scenarios...what would be an example of something that might require human intervention?} (P2)
    \end{tabular}
         \\ 
 \midrule

    \begin{tabular}{>{\centering\arraybackslash}p{1.4cm}} Examples of intended use \\(N=6)\end{tabular}&
    \begin{tabular}{p{6cm}}
    
Appreciated this section to help them jump-start divergent thinking and generating ideas.

    \end{tabular}
        &  
    \begin{tabular}{p{5cm}}
\textit{Examples of intended use was the one that weighed the most for me because looking over the examples then I can begin to extrapolate that and apply it to the problem I have} (P9)
    \end{tabular}
         
         \\ 

\bottomrule
    \end{tabular}
    \caption{\rv{Summary of} participants' comments about the \textbf{categories of information provided} in the task, ranked by the number of participants discussing each.}
    \label{tab:information}
    }
\end{table*}

\begin{table*}[h]

    \small{
\sffamily
    \centering
    \begin{tabular}{ccc}
   
    \textbf{Information Category} & \textbf{Summary} & \textbf{Example Quotes}\\
     \toprule
  \multicolumn{3}{c}{Additional information needed}\\
  \toprule
     \begin{tabular}{>{\centering\arraybackslash}p{1.5cm}} Explanation \\(N=13)\end{tabular}&
    \begin{tabular}{p{5.4cm}}
While some asked for ``explainability,'' most expressed such needs by asking the ``how'' and ``why'' questions, e.g., ``how does the model summarize?'' or ``why does it extract this sentence?''.  Also hypothesized the ``how'' from examples.

 \vspace{0.1cm}

Often interested in anticipating general patterns in model outputs or impacting factors.

 \vspace{0.1cm}

The access to explanations was considered an advantage of being able to talk to model developers directly.
    \end{tabular}
        &  
    \begin{tabular}{p{5.4cm}}
    \textit{If I knew it was looking out for, like sentences that are very definitive, then I would understand that, this isn't going to work for an opinion piece.} (P12)
    
     \vspace{0.1cm}
    \textit{I would like to learn more about the model on how it's extracting... Like there are a lot of transitional words, how does those get filtered out?... so [I know] how is the output of the model like, how easy is it for users to consume?} (P19) 
    \end{tabular}
    \\
    \midrule
    
         \begin{tabular}{>{\centering\arraybackslash}p{1.5cm}} Training data  \\(N=12)\end{tabular}&
    \begin{tabular}{p{5.4cm}}
 Interested in understanding the training data because it could help them infer impacting factors.

 \vspace{0.1cm}

 Also interested in the data shift---whether the training data matches data of their platform, to assess suitability and limitations.
    \end{tabular}
        &  
    \begin{tabular}{p{5.4cm}}

    \textit{What types of articles that thing was trained on, what diversity of articles, what type of language like formal or informal?} (P12) 
    
    \vspace{0.1cm}
    
    \textit{What it's been trained on or how it performs in relation to the types of articles being shared on this platform to evaluate that sort of match.} (P7)
    \end{tabular}
    \\
    \midrule
    
             \begin{tabular}{>{\centering\arraybackslash}p{1.5cm}}(Disaggregated) evaluation  \\(N=6)\end{tabular}&
    \begin{tabular}{p{5.4cm}}
 Mostly interested in disaggregated evaluation to understand how performance varies by impacting factors. The quantification could help them better assess their potential impact.
    \end{tabular}
        &  
    \begin{tabular}{p{5.4cm}}

    \textit{In the limitations it was highlighted a bit, but we need to quantify that... in the end, what matters is how will that impact the customer experience... if the model is 20\% accurate for topics where it's weak then I want to know that so that I can avoid summarizing for those topics} (P10)
    \end{tabular}
    \\
    \midrule

         \begin{tabular}{>{\centering\arraybackslash}p{1.5cm}} Confidence/ uncertainty  \\(N=6)\end{tabular}&
    \begin{tabular}{p{5.4cm}}
Asked whether the model could generate confidence scores. Gravitated towards using it to put guardrails on low-quality outputs. 
    \end{tabular}
        &  
    \begin{tabular}{p{5.4cm}}

    \textit{I'd want to be able to say the degree of confidence in this. I don't know... I don't think that was documented} (P7) 
    \end{tabular}
    \\
    \midrule

         \begin{tabular}{>{\centering\arraybackslash}p{1.5cm}} Customizability, improvability, roadmap \\(N=5)\end{tabular}&
    \begin{tabular}{p{5.4cm}}
Interested in knowing whether the model could be customized or improved, and whether the service-provider plans to improve it in the future, to help them plan the design accordingly and coordinate or negotiate with the team. 
    \end{tabular}
        &  
    \begin{tabular}{p{5.4cm}}

    \textit{Is anything coming up in the future?...Because if you start building for these, then some new feature gets unlocked or constraints gets erased... So I wish we had been thinking ahead for what we wanted to design} (P13) 
    \end{tabular}
    \\
    \midrule 
    
             \begin{tabular}{>{\centering\arraybackslash}p{1.5cm}} Analysis of output patterns \\(N=4)\end{tabular}&
    \begin{tabular}{p{5.4cm}}
Interested in understanding the general patterns of outputs, such as lengths and types of words.  
    \end{tabular}
        &  
    \begin{tabular}{p{5.4cm}}

    \textit{Descriptive statistics around the model output...like, is there a pattern? Does it take from the very beginning? How long is it usually? } (P2) 
    \end{tabular}
    \\
    \midrule

                 \begin{tabular}{>{\centering\arraybackslash}p{1.5cm}} Algorithm and development background\\(N=4)\end{tabular}&
    \begin{tabular}{p{5.4cm}}
Wished to understand the background of the model to infer potential biases or mismatching assumptions for their use case.

 \vspace{0.1cm}

``Technical'' knowledge could help designers build AI literacy and communicate with data scientists.
    \end{tabular}
        &  
    \begin{tabular}{p{5.4cm}}

    \textit{Understanding who, when and how it was developed. I wanna know... what are their interests? What are their biases?} (P12) 
    
     \vspace{0.1cm}
    
    \textit{The model type and algorithm because it's also about educating... so when we communicate with data scientists, we can use the same language} (P5)
    \end{tabular}
    \\
    \midrule

                 \begin{tabular}{>{\centering\arraybackslash}p{1.5cm}} Governance information\\(N=2)\end{tabular}&
    \begin{tabular}{p{5.4cm}}
Sought ``delegated trust'' by relying on their company or other organizations to vet the capabilities and ethical considerations of the model.
    \end{tabular}
        &  
    \begin{tabular}{p{5.4cm}}

    \textit{Due diligence on the service provider. What have you done as bias mitigation? Are you an ethical actor? I wanna see some sort of assurance of that, [from] a third party that I can trust.} (P9) 
    \end{tabular}
    \\
    \midrule 
    
    \end{tabular}
    \caption{\rv{Summary of }participants' comments about \textbf{additional information} participants asked for, ranked by the number of participants discussing each.}
    \label{tab:information-2}}
\end{table*}

\begin{figure*}[h]
  \centering
  \includegraphics[width=\columnwidth]{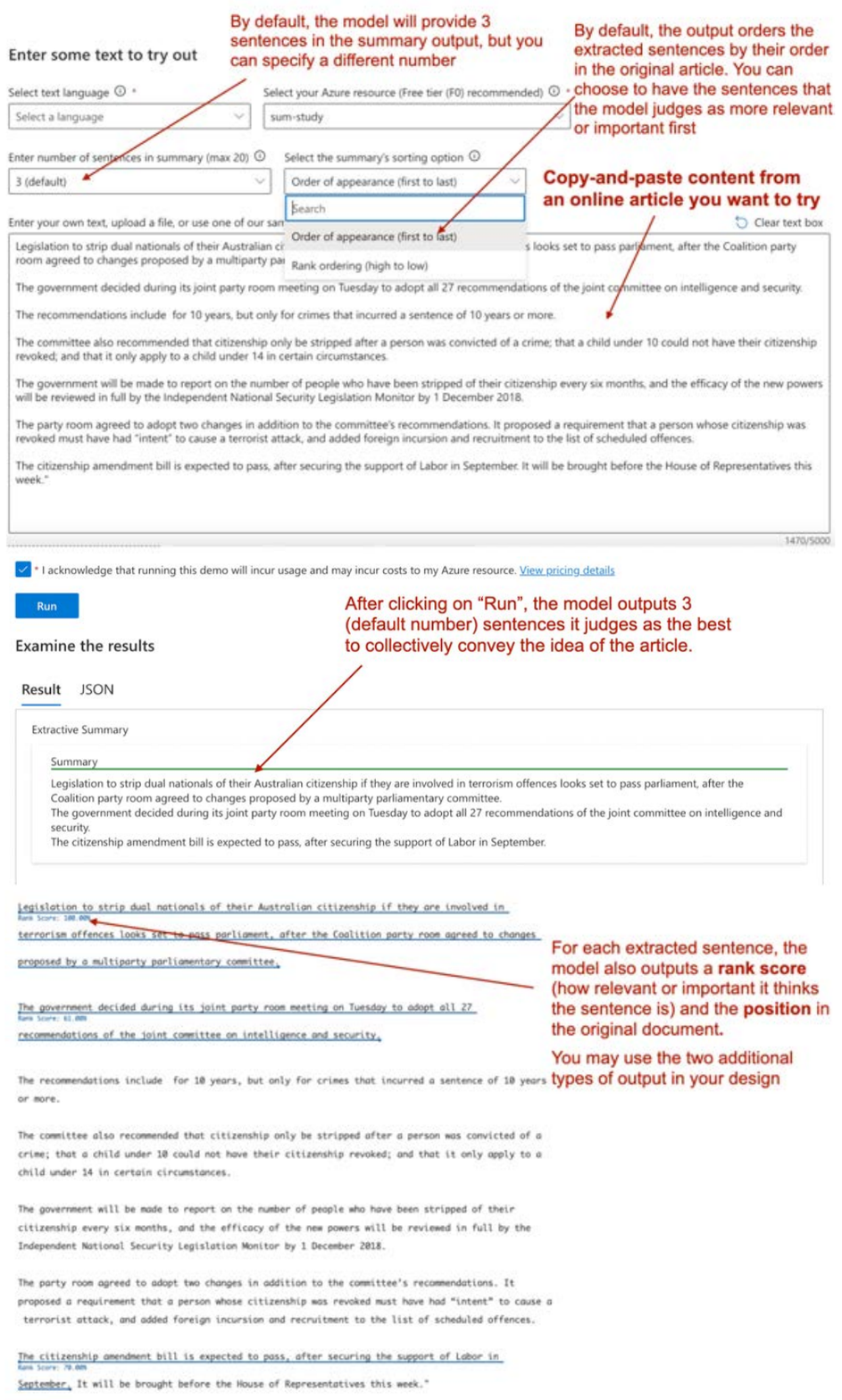}

    \vspace{-15em}

  \caption{\rv{Screenshot of the playground UI with an example input document and its model output, as retrieved in June 2022.}}~\label{fig:playgrond} 
  \Description{Screenshot showing the playground UI of the model used in the study (extractive summary provided by Microsoft Azure Cognitive Service). There is a text field to input an article, and a ``Run'' button. After clicking on the button, the output appears, showing the extractive summary for the input articles.}
    \vspace{-2.5em}
\end{figure*}

\end{document}